\documentclass[apj]{emulateapj}
\newcommand\submitms{n}		% set to y to follow AAS ``ms'' names, etc.
\if\submitms y
\else
\usepackage{apjfonts}
\fi
\usepackage{graphicx}
\usepackage{ifthen}
\usepackage{natbib}

\shorttitle{The Orbit of WASP-12b}
\shortauthors{Campo {\em et al.}}

\newcommand\degree{\degr}
\newcommand\degrees\degree

\newcounter{fignum}

\DeclareSymbolFont{UPM}{U}{eur}{m}{n}
\DeclareMathSymbol{\umu}{0}{UPM}{"16}
\let\oldumu=\umu
\renewcommand\umu{\ifmmode\oldumu\else\math{\oldumu}\fi}
\newcommand\micro{\umu}
\renewcommand\micron{\micro m}
\newcommand\microns \micron

\let\oldsim=\sim
\renewcommand\sim{\ifmmode\oldsim\else\math{\oldsim}\fi}
\let\oldpm=\pm
\renewcommand\pm{\ifmmode\oldpm\else\math{\oldpm}\fi}
\newcommand\by{\ifmmode\times\else\math{\times}\fi}

\newcommand\tablebox[1]{\begin{tabular}[t]{@{}l@{}}#1\end{tabular}}
\newbox{\wdbox}
\renewcommand\c{\setbox\wdbox=\hbox{,}\hspace{\wd\wdbox}}
\renewcommand\i{\setbox\wdbox=\hbox{i}\hspace{\wd\wdbox}}

\newcommand\herenote[1]{{\bfseries #1}\typeout{======================> note on page \arabic{page} <====================}}

\newcount\timect
\newcount\hourct
\newcount\minct
\newcommand\now{\timect=\time \divide\timect by 60
         \hourct=\timect \multiply\hourct by 60
         \minct=\time \advance\minct by -\hourct
         \number\timect:\ifnum \minct < 10 0\fi\number\minct}

\newcommand\mctc{\multicolumn{2}{c}}

\catcode`@=11

\if\submitms y
\renewcommand\comment[1]{}
\else
\newcommand\comment[1]{}
\fi
\newcommand\commenton{\catcode`\%=14}
\newcommand\commentoff{\catcode`\%=12}

\renewcommand\math[1]{$#1$}
\newcommand\mathshifton{\catcode`\$=3}
\newcommand\mathshiftoff{\catcode`\$=12}

\comment{the backslash is necessary}

\comment{alignment tab}

\let\atab=&
\newcommand\atabon{\catcode`\&=4}
\newcommand\ataboff{\catcode`\&=12}

\let\oldmsp=\sp
\let\oldmsb=\sb
\renewcommand\sp[1]{\ifmmode
	   \oldmsp{#1}%
	 \else\strut\raise.85ex\hbox{\scriptsize #1}\fi}
\renewcommand\sb[1]{\ifmmode
	   \oldmsb{#1}%
	 \else\strut\raise-.54ex\hbox{\scriptsize #1}\fi}
\newbox\@sp
\newbox\@sb
\newcommand\sbp[2]{\ifmmode%
           \oldmsb{#1}\oldmsp{#2}%
         \else
           \setbox\@sb=\hbox{\sb{#1}}%
           \setbox\@sp=\hbox{\sp{#2}}%
           \rlap{\copy\@sb}\copy\@sp
           \ifdim \wd\@sb >\wd\@sp
             \hskip -\wd\@sp \hskip \wd\@sb
           \fi
        \fi}
\newcommand\msp[1]{\ifmmode
	   \oldmsp{#1}
	 \else \math{\oldmsp{#1}}\fi}
\newcommand\msb[1]{\ifmmode
	   \oldmsb{#1}
	 \else \math{\oldmsb{#1}}\fi}
\newcommand\supon{\catcode`\^=7}
\newcommand\supoff{\catcode`\^=12}
\newcommand\subon{\catcode`\_=8}
\newcommand\suboff{\catcode`\_=12}
\newcommand\supsubon{\supon \subon}
\newcommand\supsuboff{\supoff \suboff}

\newcommand\actcharon{\catcode`\~=13}
\newcommand\actcharoff{\catcode`\~=12}

\newcommand\paramon{\catcode`\#=6}
\newcommand\paramoff{\catcode`\#=12}

\comment{And now to turn us totally on and off...}

\newcommand\reservedcharson{\commenton \mathshifton \atabon \supsubon \actcharon
	\paramon}

\newcommand\reservedcharsoff{\commentoff \mathshiftoff \ataboff
	\supsuboff \actcharoff \paramoff}

\newcommand\nojoe[1]{\reservedcharson#1\reservedcharsoff}

\catcode`@=12
\reservedcharsoff

\reservedcharson

\comment{ Must have ONLY ONE of these... trust these macros, they work

}
\reservedcharsoff

\actcharon
\if\submitms y
\newcommand\widedeltab{deluxetable}
\else
\newcommand\widedeltab{deluxetable*}
\received{2010 Mar 13}
\accepted{2010 Oct 06}
\fi

\if\submitms y
\slugcomment{\tablebox{
Submitted to {\em ApJ} 2010 March 13.}}
\else
\slugcomment{{\em ApJ}, in press.
\comment{Revision DRAFT of {\today} \now.}}
\journalinfo{}
\fi

\newcommand\ktwop{\math{k\sb{\rm{2p}}}}

\begin{document}

\title{On the Orbit of Exoplanet WASP-12b}
\author{Christopher J.\ Campo\altaffilmark{1}}
\author{Joseph Harrington\altaffilmark{1}}
\author{Ryan A.\ Hardy\altaffilmark{1}}
\author{Kevin B.\ Stevenson\altaffilmark{1}}
\author{Sarah Nymeyer\altaffilmark{1}}
\author{Darin Ragozzine\altaffilmark{2}}
\author{Nate B.\ Lust\altaffilmark{1}}
\author{David R.\ Anderson\altaffilmark{3}}
\author{Andrew Collier-Cameron\altaffilmark{4}}
\author{Jasmina Blecic\altaffilmark{1}}
\author{Christopher B.\ T.\ Britt\altaffilmark{1}}
\author{William C.\ Bowman\altaffilmark{1}}
\author{Peter J. Wheatley\altaffilmark{5}}
\author{Thomas J.\ Loredo\altaffilmark{6}}
\author{Drake Deming\altaffilmark{7}}
\author{Leslie Hebb\altaffilmark{8}}
\author{Coel Hellier\altaffilmark{3}}
\author{Pierre F.\ L.\ Maxted\altaffilmark{3}}
\author{Don Pollaco\altaffilmark{9}}
\author{Richard G.\ West\altaffilmark{10}}
\email{ccampo@knights.ucf.edu}

\affil{\sp{1}Planetary Sciences Group, Department of Physics,
 University of Central Florida, Orlando, FL 32816-2385, USA}
\affil{\sp{2}Harvard-Smithsonian Center for Astrophysics,
60 Garden St., Cambridge, MA 02138, USA}
\affil{\sp{3}Astrophysics Group, Keele University, Staffordshire ST5 
 5BG, UK}
\affil{\sp{4}School of Physics and Astronomy, University of
 St. Andrews, North Haugh, Fife KY16 9SS, UK}
\affil{\sp{5}Department of Physics, University of Warwick, Coventry,
 CV4 7AL, UK}
\affil{\sp{6}Department of Astronomy, Cornell University, Ithaca, NY 
 14853-6801, USA}
\affil{\sp{7}NASA's Goddard Space Flight Center, Greenbelt, MD
 20771-0001, USA}
\affil{\sp{8}Department of Physics and Astronomy, Vanderbilt
 University, Nashville, TN 37235, USA}
\affil{\sp{9}Astrophysics Research Centre, School of Mathematics &
Physics, Queen's University, University Road, Belfast, BT7 1NN, UK}
\affil{\sp{10}Department of Physics and Astronomy, University of
Leicester, Leicester, LE1 7RH, UK}

\begin{abstract}

We observed two secondary eclipses of the exoplanet WASP-12b using the
Infrared Array Camera on the \textit{Spitzer Space Telescope}.  The
close proximity of WASP-12b to its G-type star results in extreme
tidal forces capable of inducing apsidal precession with a period as
short as a few decades.  This precession would be measurable if the
orbit had a significant eccentricity, leading to an estimate of the
tidal Love number and an assessment of the degree of central
concentration in the planetary interior.  An initial ground-based
secondary eclipse phase reported by \citeauthor{lopez:2009} (0.510
{\pm} 0.002) implied eccentricity at the 4.5\math{\sigma} level.  The
spectroscopic orbit of \citeauthor{hebb:2009} has eccentricity 0.049
{\pm} 0.015, a 3\math{\sigma} result, implying an eclipse phase of
0.509 {\pm} 0.007.  However, there is a well documented tendency of
spectroscopic data to overestimate small eccentricities.  Our eclipse
phases are 0.5010 {\pm} 0.0006 (3.6 and 5.8 {\microns}) and 0.5006
{\pm} 0.0007 (4.5 and 8.0 {\microns}).  An unlikely orbital precession
scenario invoking an alignment of the orbit during the {\em Spitzer}
observations could have explained this apparent discrepancy, but the
final eclipse phase of \citeauthor{lopez:2010} (0.510 {\pm}
\sbp{-0.006}{+0.007}) is consistent with a circular orbit at better
than 2\math{\sigma}.  An orbit fit to all the available transit,
eclipse, and radial-velocity data indicates precession at
\math{<1\sigma}; a non-precessing solution fits better.  We
also comment on analysis and reporting for {\em Spitzer} exoplanet
data in light of recent re-analyses.

\if\submitms y
\else
\comment{\hfill\herenote{DRAFT of {\today} \now}.}
\fi
\end{abstract}
\object{WASP-12b}

\section{INTRODUCTION}
\label{intro}

When exoplanets transit (pass in front of) their parent stars as
viewed from Earth, one can constrain their sizes, masses, and orbits
\citep{charbonneau:2007, winn:2009}.  Most transiting planets also
pass behind their stars (secondary eclipse).  This allows atmospheric
characterization by measurement of planetary flux and constrains
orbital eccentricity, \math{e}, through timing and duration of the
eclipse \citep{kallrath:1999}.

WASP-12b is one of the hottest transiting exoplanets discovered to
date, with an equilibrium temperature of 2516 K for zero albedo and
uniform redistribution of incident flux \citep{hebb:2009}.  It also
has a 1.09-day period, making it one of the shortest-period transiting
planets.  The close proximity to its host star \citep[0.0229 {\pm}
  0.0008 AU,][]{hebb:2009} should induce large tidal bulges on the
planet's surface.  Tidal evolution should quickly circularize such
close-in orbits \citep{mardling:2007}.  \citet{hebb:2009} calculate a
circularization time for WASP-12b as short as 3 Myr, much shorter than
the estimated 2 Gyr age of WASP-12 or even the circularization times
estimated for other hot Jupiters, given similar planetary tidal
dissipation, though this calculation was based on a formalism
\citep{goldreich:1966} that ignores the influence of stellar tides and
the coupling of eccentricity and semi-major axis in the evolution of
the system.  The influence of stellar tides could prolong the
dissipation timescale to well over the age of the system
\citep{jackson:2008}.  The non-Keplerian gravitational potential may
cause apsidal precession, measurable as secondary eclipse and transit
timing variations over short time scales.  WASP-12b also has an
abnormally large radius (\math{R\sb{\rm{p}}} = 1.79 {\pm} 0.09 Jupiter
radii, \math{R\sb{\rm{J}}}, \citealp{hebb:2009}) compared to those
predicted by theoretical models \citep{bodenheimer:2003, fortney:2007}
and to other short-period planets.  Tidal heating models assume
non-zero \math{e}, and the heating rate can differ substantially for
different values of \math{e}.  WASP-12b's inflated radius may result
from tidal heating, but this is difficult to justify if the orbit is
circular \citep{li:2010}.

Ground-based observations by \citet{lopez:2009} detected a
secondary-eclipse phase for WASP-12b of 0.510 {\pm} 0.002, implying an
eccentric orbit at the 4.5\math{\sigma} level (\citealp{lopez:2010}
revised the uncertainty to \sbp{-0.006}{+0.007}).  Radial velocity data
\citep{hebb:2009} find \math{e} = 0.049 {\pm} 0.015, a 3\math{\sigma}
eccentricity, and predict an eclipse phase of 0.509 {\pm} 0.007.
Given an eccentric orbit and the fast predicted precession time scale,
WASP-12b makes an excellent candidate for the first direct detection
of exoplanetary apsidal precession.  Such precession has been detected
many times for eclipsing binary stars \citep{kreiner:2001}.

Against an orbit established by transit timings, precession would be
apparent in just two eclipses, if sufficiently separated in time.  For
eccentric orbits, the eclipse-transit interval can differ from the
transit-eclipse interval, and for precessing orbits this difference
varies sinusoidally over one precession period.  If the difference is
insignificant, it places an upper limit on \math{e \cos \omega}, where
\math{\omega} is the argument of periapsis.  In the case of WASP-12b,
which is expected to precess at a rate of 0.05{\degrees} d\sp{-1}
\citep{ragozzine:2009}, if the orbit is observed when \math{\omega
  \sim \pm 90 \degrees} and the effect on the eclipse timing is
maxmized, and assuming a timing precision of 0.0007 days, then
secondary eclipse observations situated five months apart could detect
precession at the 3\math{\sigma} level (see Equation
\ref{eq:gimenmod}).  We note that the method of
\citealp{batygin:2009}, based on the work of \citealp{mardling:2007}
and extended to the three-dimensional case by \citealp{mardling:2010},
is an indirect assessment of apsidal precession, since no orbital
motion is actually observed.  The technique, which only applies to
multi-planet systems with a tidally affected inner planet and a
nearby, eccentric, outer planet, cannot currently be applied to
WASP-12b.

Paired with the \citeauthor{lopez:2009} data, our \textit{Spitzer
  Space Telescope}\/ \citep{werner:2004} eclipse observations provide
a one-year baseline.  \textit{Spitzer's}\/ high photometric precision
also allows an accurate assessment of \math{e \cos \omega}.  One can
solve for \math{e} and \math{\omega} separately given \math{e \sin
  \omega} from precise radial velocity data.  The following sections
present our observations; photometric analysis; a dynamical model that
considers parameters from this work, the original and revised
parameters of \citeauthor{lopez:2009}, \citeauthor{hebb:2009}, new
transit times from the Wide-Angle Search for Planets (WASP), and
transit times from a network of amateur astronomers; and our
conclusions.

\if\submitms y
\clearpage
\fi
\begin{figure}[htb]
\if\submitms y \setcounter{fignum}{\value{figure}}
\addtocounter{fignum}{1} \newcommand\fignam{f\arabic{fignum}.eps}
\else \newcommand\fignam{figs/wa012bs41_preflash.ps} \fi
\includegraphics[width=\columnwidth, clip]{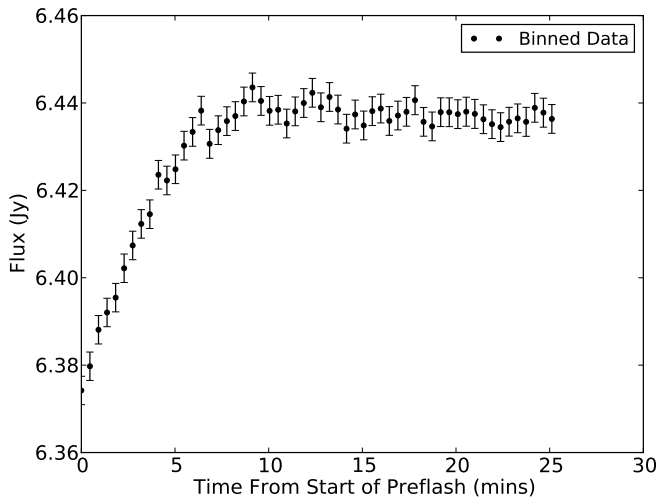}
\figcaption{\label{fig:preflash} Preflash light curve.  These are
  channel-4 (8 {\microns} data, analyzed with aperture photometry at
  the pixel location of the eclipse observations.  The preflash source
  is bright compared to WASP-12, which allows the array sensitivity to
  ``ramp'' up before the science observations.  Without a preflash,
  similar observations generally show a steeper and longer ramp in the
  eclipse observations.}
\end{figure}
\if\submitms y
\clearpage
\fi

\if\submitms y
\clearpage
\fi
\begin{figure*}[thb]
\if\submitms y
  \setcounter{fignum}{\value{figure}}
  \addtocounter{fignum}{1}
  \newcommand\fignama{f\arabic{fignum}a.eps}
  \newcommand\fignamb{f\arabic{fignum}b.eps}
  \newcommand\fignamc{f\arabic{fignum}c.eps}
\else
  \newcommand\fignama{figs/WASP-12b-raw.ps}
  \newcommand\fignamb{figs/WASP-12b-bin.ps}
  \newcommand\fignamc{figs/WASP-12b-norm.ps}
\fi
\strut\hfill
\includegraphics[width=0.32\textwidth, clip]{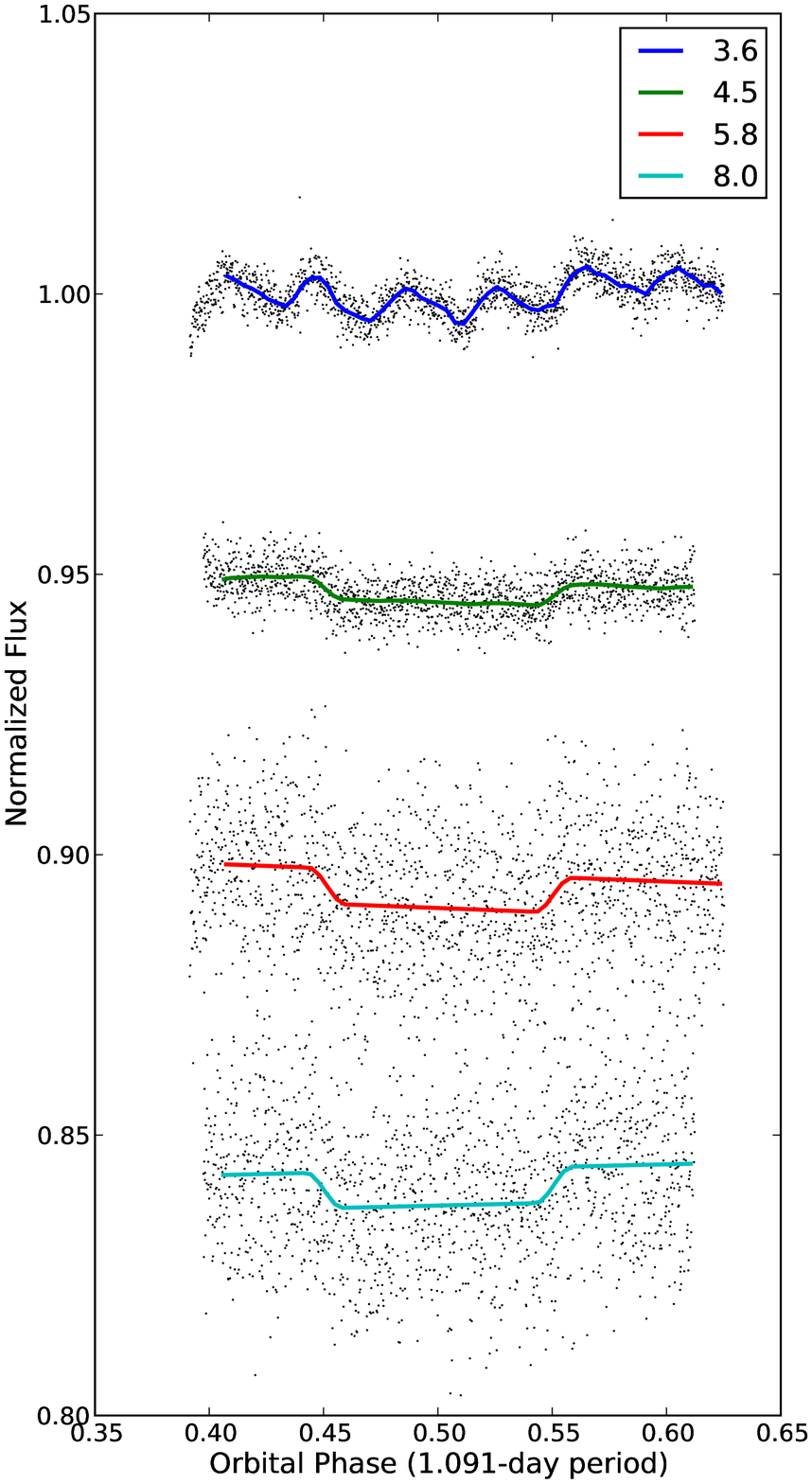}
\includegraphics[width=0.32\textwidth, clip]{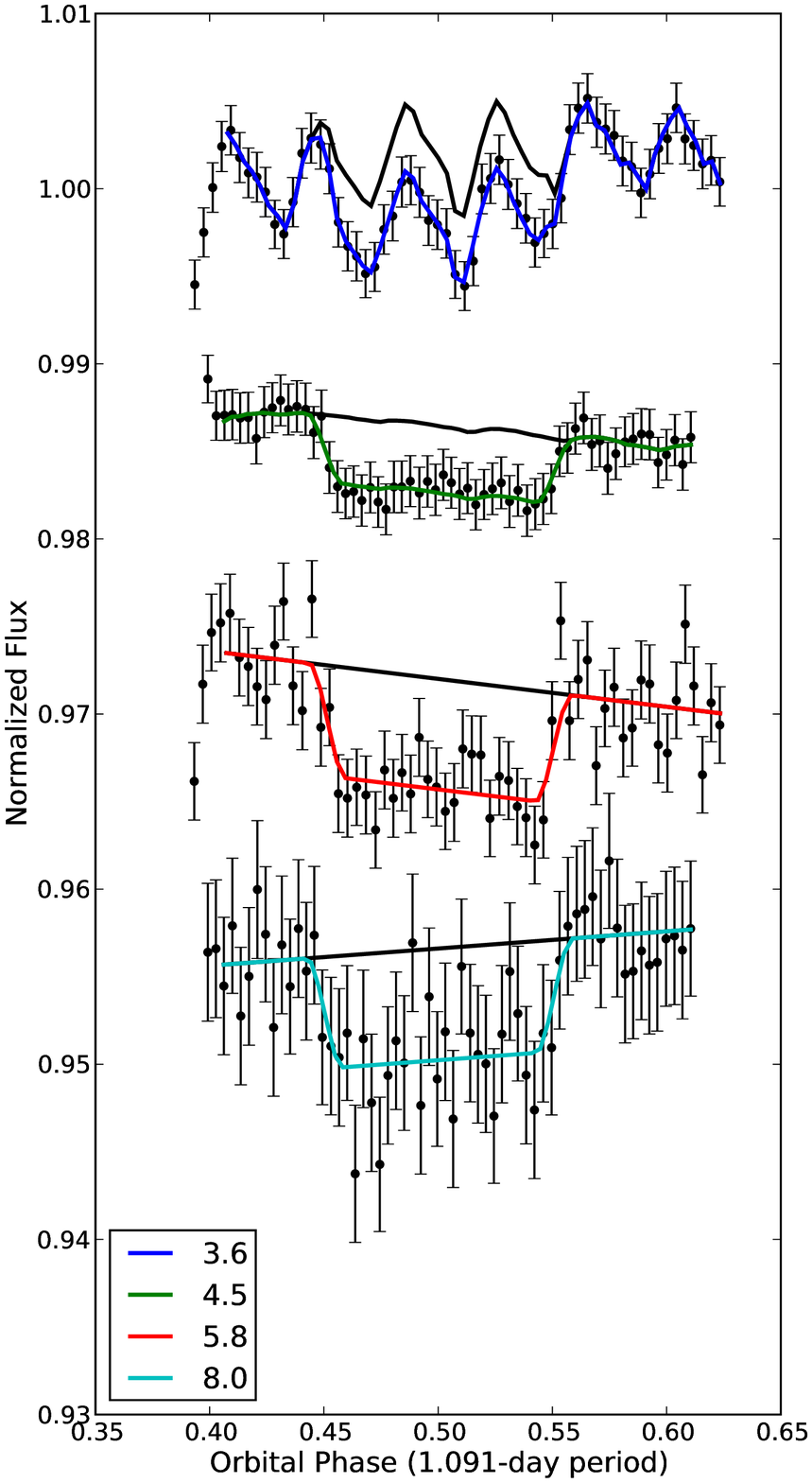}
\includegraphics[width=0.32\textwidth, clip]{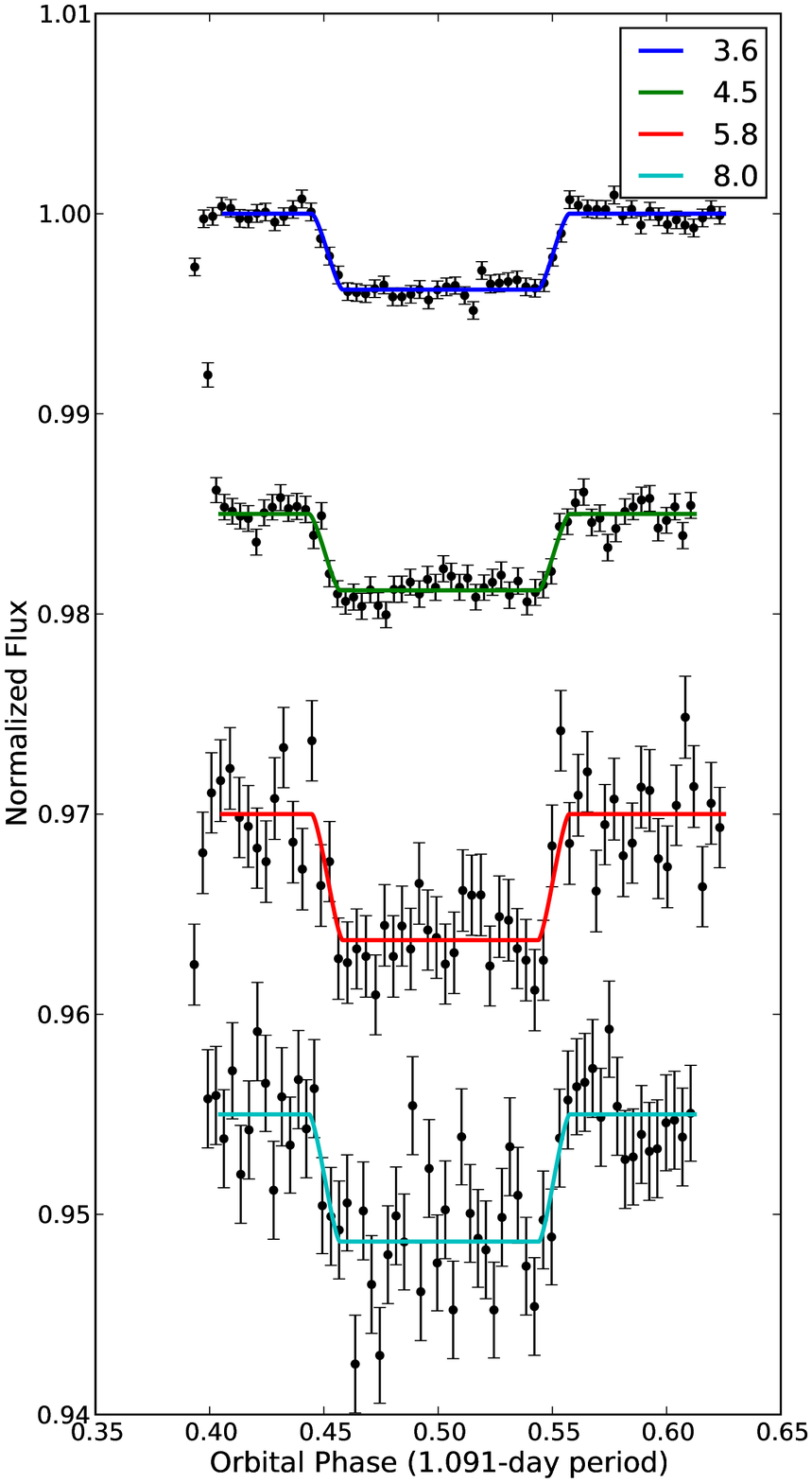}
\hfill\strut
\figcaption{\label{fig:lightcurves}
Raw (left), binned (center), and systematics-corrected (right)
secondary eclipse light curves of WASP-12b in the four IRAC channels,
normalized to the mean system flux within the fitted data.  Colored
lines are the best-fit models; black curves omit their eclipse model
elements. A few initial points in all channels are not fit, as 
indicated, to allow the telescope pointing and instrument to
stabilize.
}
\end{figure*}
\if\submitms y
\clearpage
\fi

\if\submitms y
\clearpage
\fi
\begin{figure}[thb]
\if\submitms y
  \setcounter{fignum}{\value{figure}}
  \addtocounter{fignum}{1}
  \newcommand\fignama{f\arabic{fignum}a.eps}
  \newcommand\fignamb{f\arabic{fignum}b.eps}
  \newcommand\fignamc{f\arabic{fignum}c.eps}
  \newcommand\fignamd{f\arabic{fignum}d.eps}
\else
  \newcommand\fignama{figs/wa012bs11-noisecorr.ps}
  \newcommand\fignamb{figs/wa012bs21-noisecorr.ps}
  \newcommand\fignamc{figs/wa012bs31-noisecorr.ps}
  \newcommand\fignamd{figs/wa012bs41-noisecorr.ps}
\fi
\includegraphics[width=0.22\textwidth, clip]{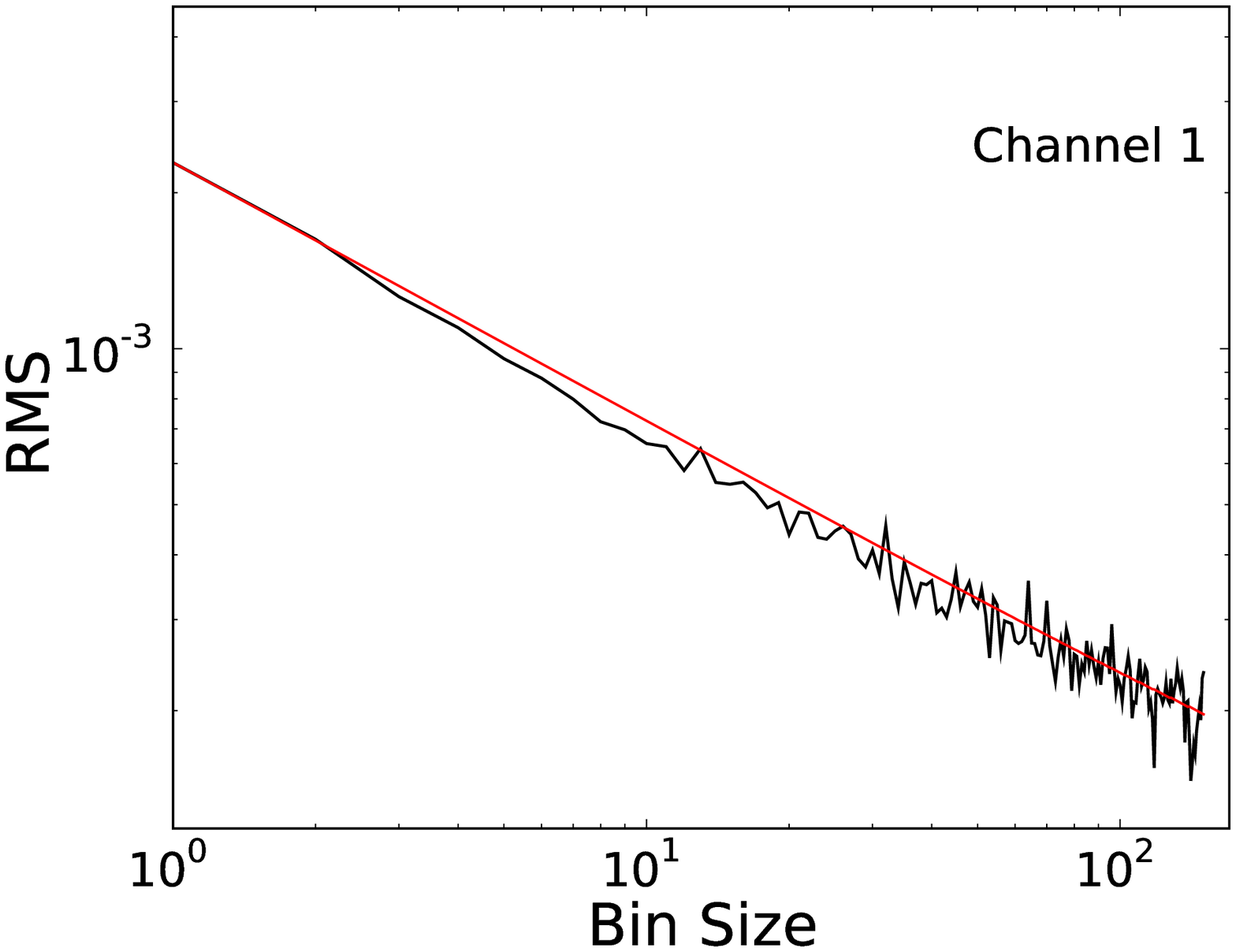}
\hfill
\includegraphics[width=0.22\textwidth, clip]{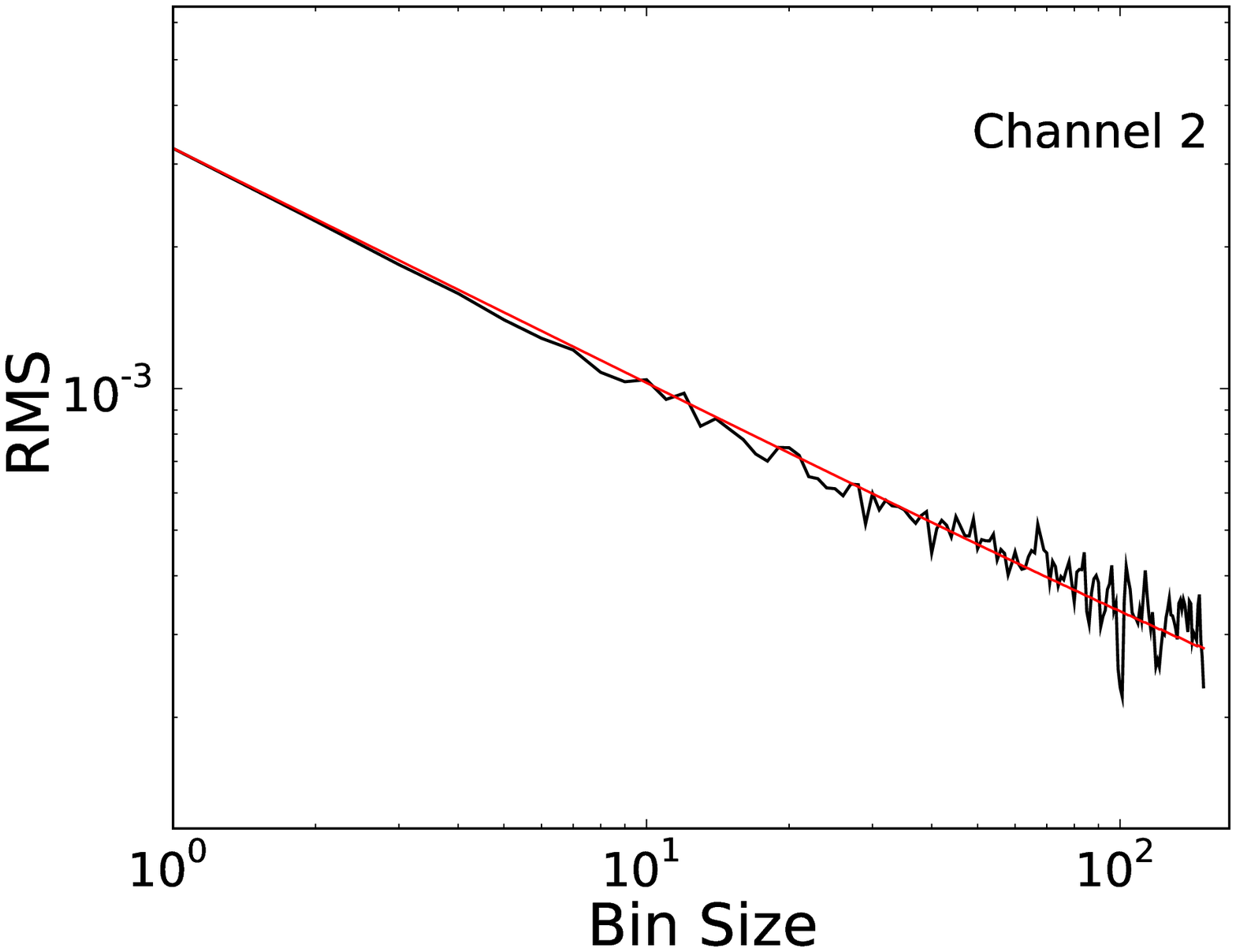}
\newline
\strut\newline
\includegraphics[width=0.22\textwidth, clip]{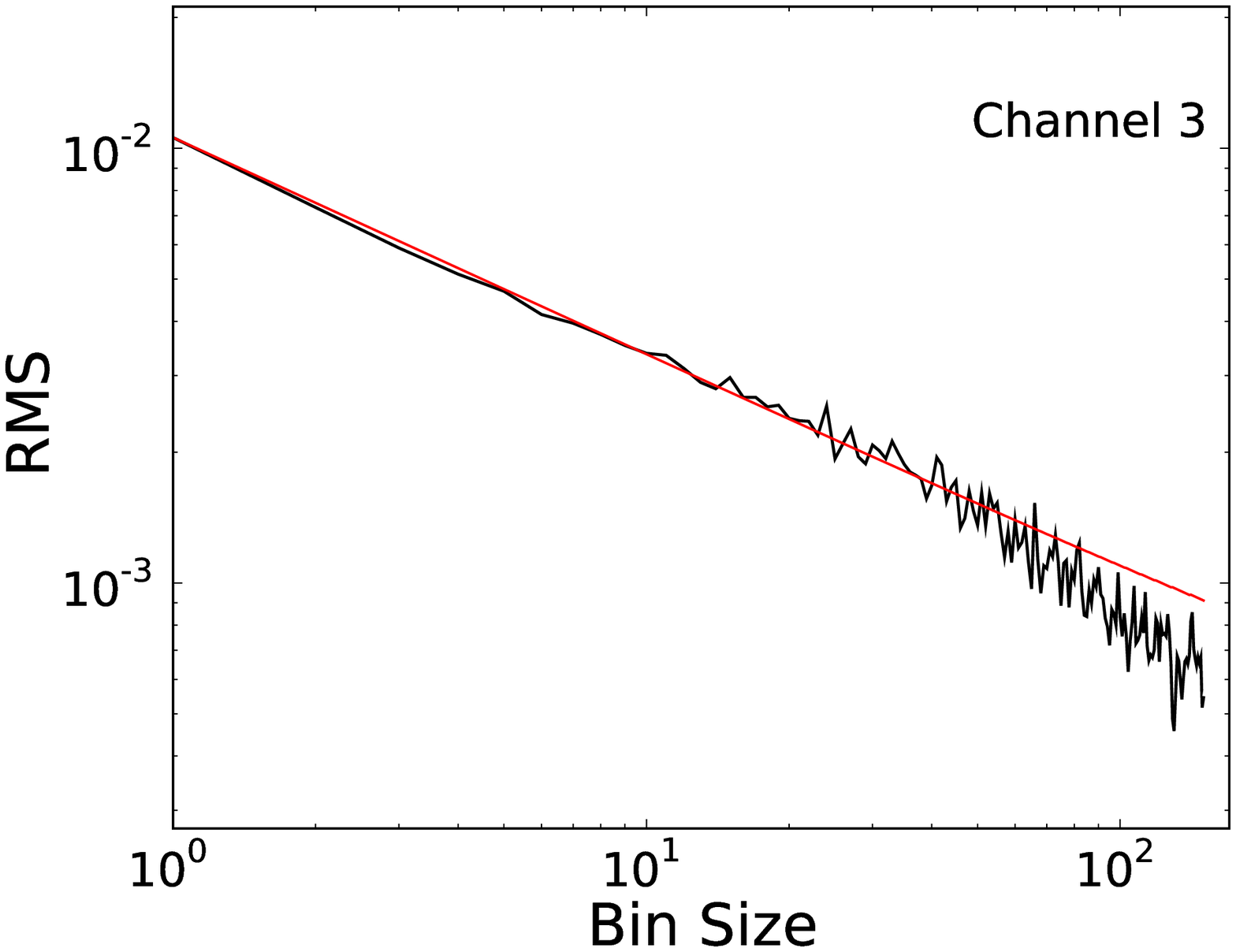}
\hfill
\includegraphics[width=0.22\textwidth, clip]{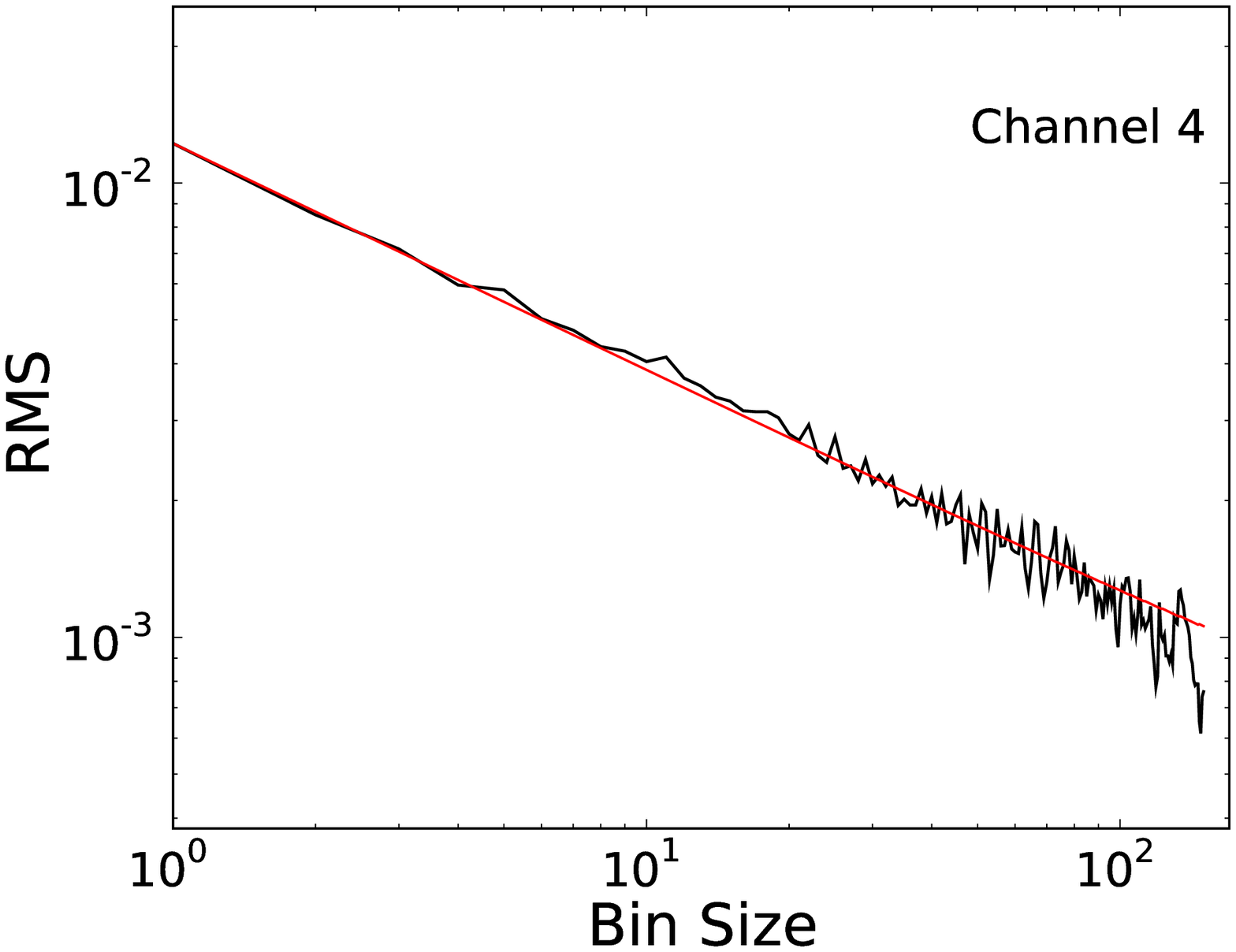}
\figcaption{\label{fig:noisecorr}
Root-mean-squared (RMS) residual flux \textit{vs.}\/ bin size in each
channel.  This plot tests for correlated noise.  The straight line is
the prediction for Gaussian white noise.  Since the data do not
deviate far from the line, the effect of correlated noise is minimal.}
\end{figure}
\if\submitms y
\clearpage
\fi

\section{OBSERVATIONS}
\label{sec:obs}

We observed two secondary eclipses of WASP-12b with the
\textit{Spitzer} Infrared Array Camera (IRAC,
\citealp{fazio:2004}) in full-array mode.  Observations on 2008
October 29 at 4.5 and 8.0 {\microns} (IRAC channels 2 and 4,
respectively) lasted 338 minutes (program ID 50759); those on 2008
November 3 at 3.6 and 5.8 {\microns} (channels 1 and 3, respectively)
lasted 368 minutes (Program ID 50517).  The IRAC beam splitter enabled
simultaneous observations in the paired channels; all exposures were
12 seconds, resulting in 1696 frames in each of channels 1 and 3 and
1549 frames in each of channels 2 and 4.  To minimize inter-pixel
variability in all channels and the known intra-pixel variability in
channels 1 and 2
\citep{reach:2005,charbonneau:2005,harrington:2007,stevenson:2010},
each target had fixed pointing.  Prior to the science observations in
channels 2 and 4, we observed a 57-frame preflash, exposing the array
to a relatively bright source to reduce the time-dependent sensitivity
(``ramp'') effect in channel 4
(\citealp{charbonneau:2005,harrington:2007,knutson:2008}, see Figure
\ref{fig:preflash}).  Each observation ended with a 10-frame,
post-eclipse observation of blank sky in the same array positon as the
science observations to check for warm pixels in the photometric
aperture.

\section{DATA ANALYSIS}
\label{sec:anal}

\textit{Spitzer's}\/ data pipeline (version S18.7.0) applied both
standard and IRAC-specific corrections, producing the Basic Calibrated
Data (BCD) we analyzed.  Our analysis pipeline masks pixels according
to \textit{Spitzer's} permanent bad pixel masks.  It masks additional
bad pixels (e.g., from cosmic-ray strikes), by grouping frames into
sets of 64 and doing a two-iteration outlier rejection at each pixel
location.  Within each array position in each set, this routine
calculates the standard deviation from the \textit{median}, masks any
pixels with greater than \math{4\sigma} deviation, and repeats this
procedure once.  Masked pixels do not participate in the analysis.

The channel-4 data show a horizontal streak of pixels with low fluxes
located \sim10 pixels above the star.  A similar diagonal streak
appears \sim10 pixels below and left of the star.  This artifact,
which we masked, resulted from saturation in a prior observation.  A
two-dimensional Gaussian fit found the photometry center for each
image \citep[see the Supplementary Information for discussion of
  centering methods on {\em Spitzer}\/ data]{stevenson:2010}.  The
pipeline uses interpolated aperture photometry
\citep{harrington:2007}, ignoring frames with masked pixels in the
photometry aperture and not using masked pixels in sky level averages.
Table \ref{tab:fits} presents photometry parameters.  We evaluated
numerous photometry apertures (see Table \ref{tab:ramp} in the
appendix), choosing the one with the best final light-curve fit in
each channel (see below).  Because channel 4 had a higher background
flux level, the best sky annulus was larger and the photometry
aperture was smaller than in the other channels.  The channel-4
aperture contained 63% of the point-spread function; the others
contained 89% or more.

The intra-pixel variation only affects channels 1 and 2, and was only
substantial in channel 1 (see Table \ref{tab:fits} and Figure
\ref{fig:lightcurves}).  We model the intra-pixel effect with a
second-order, two-dimensional polynomial,
\begin{eqnarray}
\label{eq:vip}
V\sb{IP}(x,y) = p\sb1y\sp2 + p\sb2x\sp2 + p\sb3xy + p\sb4y + p\sb5x + 1,
\end{eqnarray}
where \math{x} and \math{y} are the centroid coordinates relative to
the pixel center nearest the median position and \math{p\sb{1}},
\math{p\sb{2}}, \math{p\sb{3}}, \math{p\sb{4}}, and \math{p\sb{5}} can
be free parameters.  We model the ramp for channel 1 with the rising
exponential
\begin{eqnarray}
R(t) = 1 - \exp \left( -r\sb{1}[t - r\sb{2}] \right),
\end{eqnarray}
where \math{t} is orbital phase and \math{r\sb{1}} and \math{r\sb{2}}
are free parameters.  The remaining channels used a linear model,
\begin{eqnarray}
\label{eq:linramp}
R(t) = r\sb{3}(t - 0.5) + 1,
\end{eqnarray}
where \math{r\sb{3}} is a free parameter.  The eclipse, \math{E(t)},
is a \citet{mandelagol:2002} model, assuming no limb darkening.  The
final light-curve model is
\begin{eqnarray}
\label{eq:lcmodel}
F(x, y, t) = F\sb{s}V\sb{IP}(x, y)R(t)E(t),
\end{eqnarray}
where \math{F(x, y, t)} is the flux measured from interpolated
aperture photometry and \math{F\sb{s}} is the (constant) system flux
outside of eclipse, including the planet.

\if\submitms y
\clearpage
\fi
\atabon\begin{\widedeltab}{rr@{\,{\pm}\,}rr@{\,{\pm}\,}rr@{\,{\pm}\,}rr@{\,{\pm}\,}r}
\if\submitms y
\tabletypesize{\scriptsize}
\fi
\tablecaption{\label{tab:fits} Joint Light-Curve Fit Parameters}
\tablewidth{0pt}
\tablehead{
\colhead{Parameter} &
\mctc{3.6 {\microns}} &
\mctc{4.5 {\microns}} &
\mctc{5.7 {\microns}} &
\mctc{8.0 {\microns}}
}
\startdata
Array Position (\math{\bar{x}}, pix)                               &          \mctc{25.20}       &          \mctc{20.24}       &       \mctc{19.35}          &       \mctc{21.45}        \\
Array Position (\math{\bar{y}}, pix)                               &       \mctc{26.98}          &       \mctc{27.95}          &       \mctc{27.15}          &       \mctc{25.67}        \\
Position Consistency\tablenotemark{a} (\math{\delta\sb{x}}, pix)   &        \mctc{0.012}         &        \mctc{0.013}         &        \mctc{0.030}         &        \mctc{0.13}        \\
Position Consistency\tablenotemark{a} (\math{\delta\sb{y}}, pix)   &        \mctc{0.012}         &        \mctc{0.013}         &        \mctc{0.018}         &        \mctc{0.14}        \\
Aperture Size (pix)                                                &          \mctc{3.75}        &          \mctc{4.00}        &          \mctc{2.75}        &          \mctc{2.00}      \\
Sky Annulus Inner Radius (pix)                                     &          \mctc{7.00}        &          \mctc{7.00}        &          \mctc{7.00}        &          \mctc{12.00}     \\
Sky Annulus Outer Radius (pix)                                     &          \mctc{12.00}       &          \mctc{12.00}       &          \mctc{12.00}       &          \mctc{30.00}     \\
System Flux \math{F\sb{s}} (\micro Jy)                             &         25922 &  11         &      16614    &  3          &       11129   &  4          &       6111   &  3         \\
Eclipse Depth\tablenotemark{c} (\math{F\sb{\rm p}/F\sb{\rm *}})    &       0.00379 &  0.00013    &       0.00382 &  0.00019    &       0.00629 &  0.00052    &      0.00636 &  0.00067   \\
Brightness Temperature (K)                                         &          2740 &  49         &          2571 &  73         &          3073 &  176        &        2948   &  233      \\
Eclipse Mid-time\tablenotemark{b, c} (\math{t\sb{\rm mid}}, phase) &        0.5010 &  0.0006     &        0.5006 &  0.0007     &        0.5010 &  0.0006     &        0.5006 &  0.0007   \\
Eclipse Mid-time\tablenotemark{d} (\math{t\sb{\rm mid}}, BJD - 2,454,000) & 773.6481 &  0.0006   &       769.2819 &  0.0008    &        773.6481 &  0.0006   &       769.2819 &  0.0008  \\
Eclipse Duration\tablenotemark{c} (\math{t\sb{\rm 4-1}}, sec)      &      10615.66 & 102.95      &      10749.97 & 142.72      &      10615.66 & 102.95      &      10749.97 & 142.72    \\
Ingress Time (\math{t\sb{\rm 2-1}}, sec)                           &        \mctc{1266.43}       &        \mctc{1266.43}       &        \mctc{1266.43}       &        \mctc{1266.43}     \\
Egress Time (\math{t\sb{\rm 4-3}}, sec)                            &        \mctc{1266.43}       &        \mctc{1266.43}       &        \mctc{1266.43}       &        \mctc{1266.43}     \\
Ramp Name                                                          &  \mctc{Rising Exponential}  &         \mctc{Linear}       &      \mctc{Linear}          &     \mctc{Linear}         \\
Ramp, Curvature\tablenotemark{c} (\math{r\sb{1}})                  &      29      &  1           &           \mctc{0}          &           \mctc{0}          &          \mctc{0}         \\
Ramp, Phase Offset\tablenotemark{c} (\math{r\sb{2}})               &      \mctc{0.17747}         &           \mctc{0.5}        &           \mctc{0.5}        &           \mctc{0.5}      \\
Ramp, Linear Term\tablenotemark{c} (\math{r\sb{3}})                &           \mctc{0}          &      -0.0102  &  0.0015     &      -0.016  &  0.004       &        0.010   &  0.005   \\
Intra-pixel, Quadratic Term in y\tablenotemark{c} (\math{p\sb{1}}) &           \mctc{0}          &      -0.09   &  0.04        &           \mctc{0}          &          \mctc{0}         \\
Intra-pixel, Quadratic Term in x\tablenotemark{c} (\math{p\sb{2}}) &      -0.140   &  0.011      &           \mctc{0}          &           \mctc{0}          &          \mctc{0}         \\
Intra-pixel, Cross Term (\math{p\sb{3}})                           &           \mctc{0}          &           \mctc{0}          &           \mctc{0}          &          \mctc{0}         \\
Intra-pixel, Linear Term in y\tablenotemark{c} (\math{p\sb{4}})    &       0.086   &  0.004      &           \mctc{0}          &           \mctc{0}          &          \mctc{0}         \\
Intra-pixel, Linear Term in x (\math{p\sb{5}})                     &           \mctc{0}          &           \mctc{0}          &           \mctc{0}          &          \mctc{0}         \\
\comment{
Intra-pixel, Constant Term (\math{p\sb{6}})                        &           \mctc{1.0}        &           \mctc{1.0}        &           \mctc{0}          &          \mctc{0}         \\}
Total frames                                                       &         \mctc{1697}         &         \mctc{1560}         &         \mctc{1697}         &         \mctc{1560}       \\
Good frames\tablenotemark{e}                                       &         \mctc{1532}         &         \mctc{1457}         &         \mctc{1543}         &         \mctc{1467}       \\
Rejected frames\tablenotemark{e} (%)                               &         \mctc{9}            &         \mctc{6}            &         \mctc{9}            &         \mctc{5}          \\
Free Parameters                                                    &          \mctc{10}          &           \mctc{9}          &           \mctc{10}         &           \mctc{9}        \\
Number of Data Points in Fit                                       &         \mctc{3075}         &         \mctc{2924}         &         \mctc{3075}         &         \mctc{2924}       \\
BIC                                                                &      \mctc{3155.5}          &          \mctc{2996.0}      &          \mctc{3155.5}      &          \mctc{2996.0}    \\
AIC                                                                &      \mctc{3095.7}          &          \mctc{2942.2}      &          \mctc{3095.7}      &          \mctc{2942.2}    \\
Standard Deviation of Normalized Residuals                         &      \mctc{0.00228716}      &      \mctc{0.00324027}      &      \mctc{0.01058880}      &      \mctc{0.01222100}    \\
Uncertainty Scaling Factor                                         &      \mctc{0.31248}         &      \mctc{0.44500}         &      \mctc{0.91832}         &      \mctc{0.62475}       \\

\enddata
\tablenotetext{a}{RMS frame-to-frame position difference.}
\tablenotetext{b}{Based on the transit ephemeris time given by
  \citet{hebb:2009}.}
\tablenotetext{c}{MCMC jump parameter.}
\tablenotetext{d}{Uncorrected for light-travel time in the
  exoplanetary system (see Dynamics section).}
\tablenotetext{e}{We reject frames during instrument/telescope
  settling and with bad pixels in the photometry aperture.}
\end{\widedeltab}\ataboff
\if\submitms y
\clearpage
\fi
\placetable{tab:fits}

To estimate photometric uncertainties, we propagate the values in the
\textit{Spitzer}\/ BCD uncertainty images through the aperture
photometry calculation.  Since the \textit{Spitzer} pipeline generally
overestimates uncertainties, we fit an initial model with a
\math{\chi\sp{2}} minimizer and then scale all uncertainties to give a
reduced \math{\chi\sp{2}} of unity \citep{harrington:2007}.  We
confirm the fit by redoing it with the new uncertainties.  The scaling
factor is proportional to the standard deviation of the normalized
residuals (SDNR) from the models, as reported in Tables \ref{tab:ramp}
(in the appendix) and \ref{tab:fits}.  The \sim 2% SDNR variation does
not significantly affect the fits.  To select among models, we must
compare fits made to the same data, including uncertainties.  So, we
use just one uncertainty scaling factor for all models in each
combination of aperture and channel (see Tables \ref{tab:fits} and
\ref{tab:ramp} in the appendix).

\citet{sivia:2006} provide an accessible tutorial to the Bayesian
approach of our subsequent analysis.  \citet[chapter 29, and
  especially section 4]{mackay:2003} introduces Markov-Chain Monte
Carlo (MCMC) and discusses its practicalities.  Briefly, the MCMC
algorithm calculates \math{\chi\sp{2}} at random locations near the
\math{\chi\sp{2}} minimum in the parameter phase space, accepting only
some of these steps for later analysis.  The density of these accepted
points is proportional to the probability of a model at that location,
given the data.  The attraction of MCMC is that histograms and scatter
plots of subsets of interesting parameters from the accepted points
display parameter uncertainties and correlations in a way that fully
accounts for the uncertainties in and correlations with the
uninteresting parameters.  These are called marginal distributions.

\begin{figure}[htb]
\if\submitms y \setcounter{fignum}{\value{figure}}
\addtocounter{fignum}{1} \newcommand\fignam{f\arabic{fignum}.eps}
\else \newcommand\fignam{figs/ch13-hist.eps} \fi
\includegraphics[width=\columnwidth, clip]{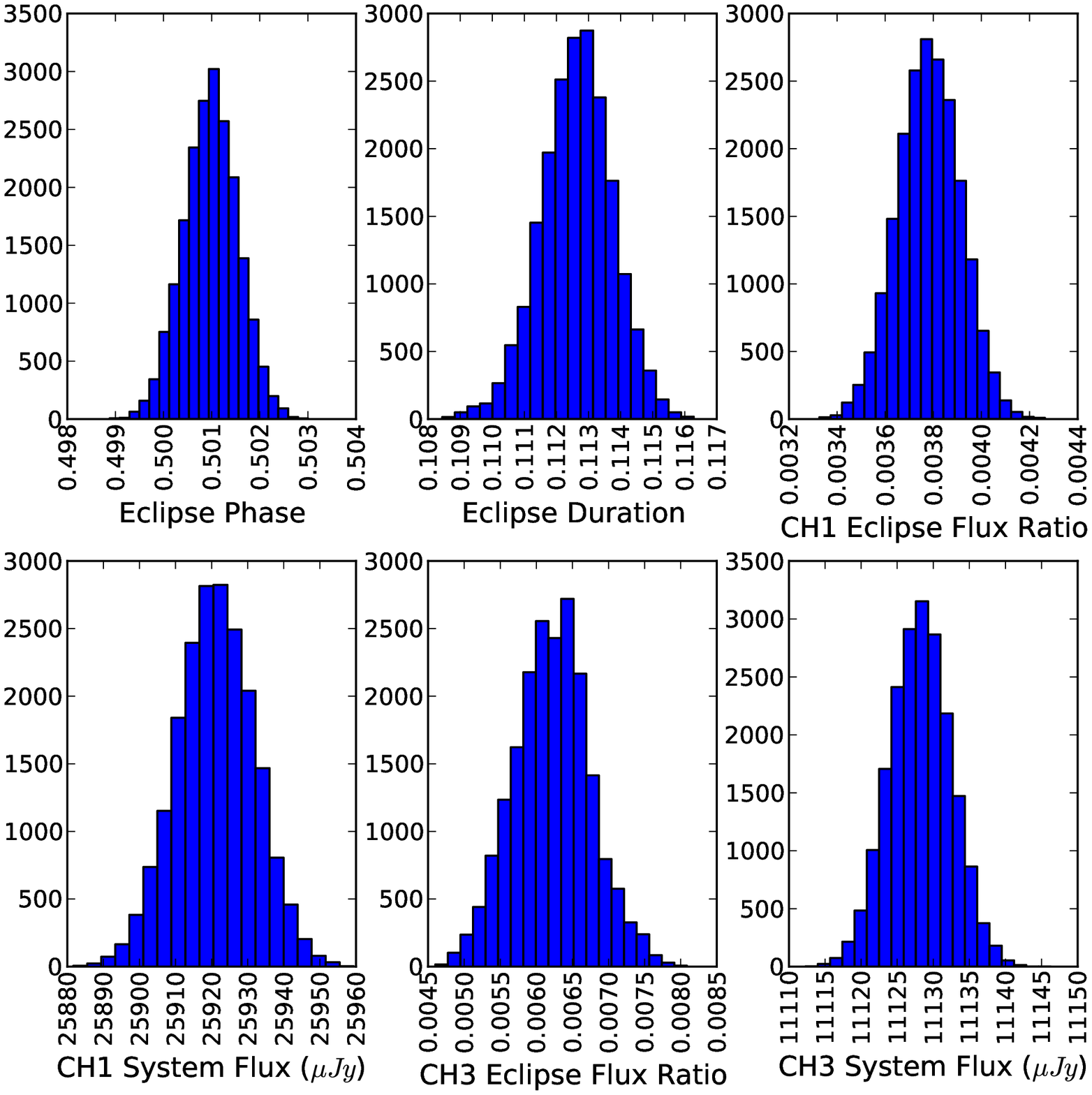}
\figcaption{\label{fig:hist13} Parameter histograms for 3.6 and 5.8
  {\microns}.  To decorrelate the Markov chains, the histograms come
  from every 100\sp{th} MCMC step.}
\end{figure}

\begin{figure}[htb]
\if\submitms y \setcounter{fignum}{\value{figure}}
\addtocounter{fignum}{1} \newcommand\fignam{f\arabic{fignum}.eps}
\else \newcommand\fignam{figs/ch24-hist.eps} \fi
\includegraphics[width=\columnwidth, clip]{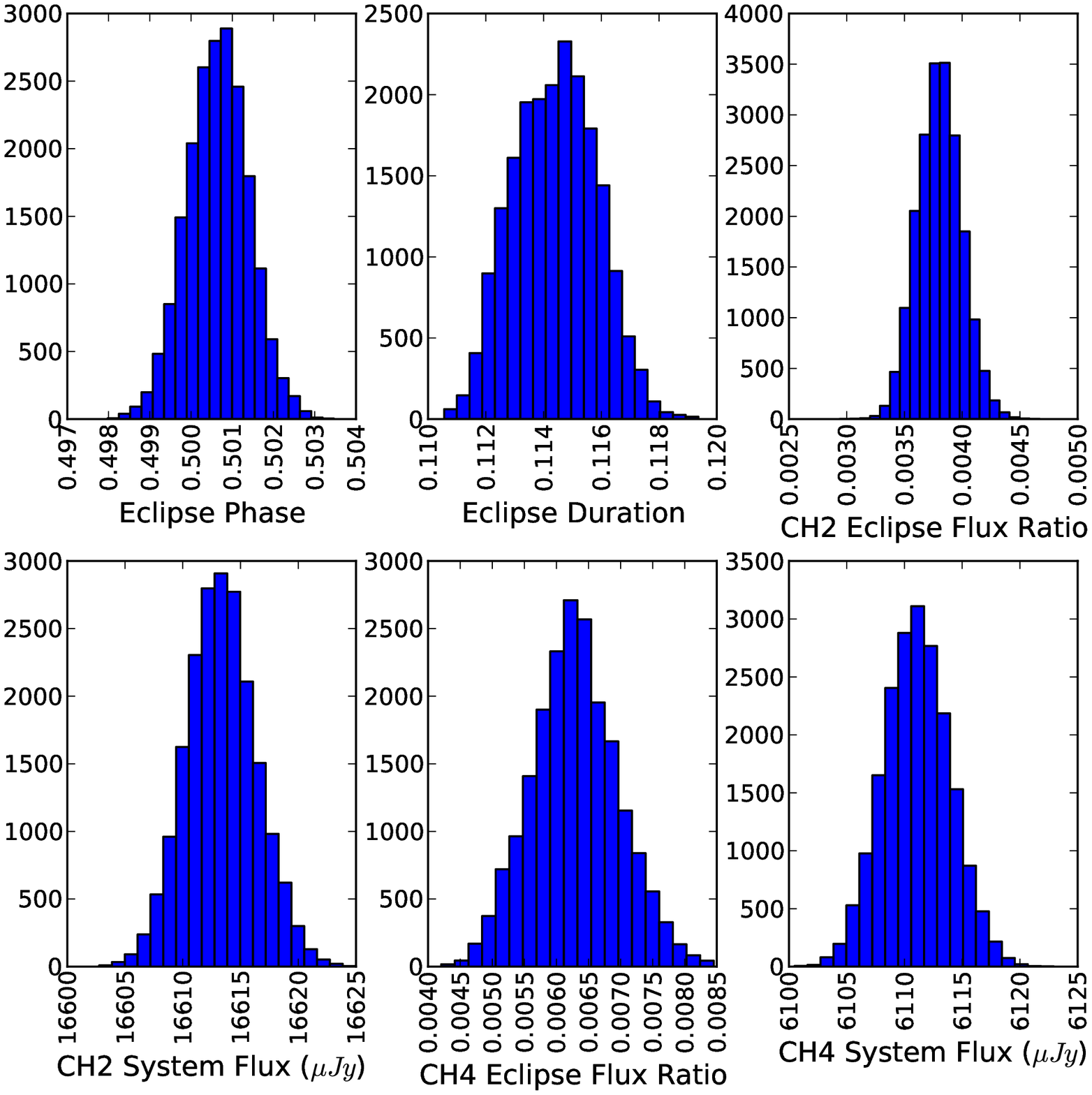}
\figcaption{\label{fig:hist24} Parameter histograms for 4.5 and 8.0
  {\microns}.  To decorrelate the Markov chains, the histograms come
  from every 100\sp{th} MCMC step.}
\end{figure}

We fit equation \ref{eq:lcmodel} with a \math{\chi\sp{2}} minimizer
and assess parameter uncertainties with a Metropolis random-walk (MRW)
MCMC algorithm.  Our MRW used independent Gaussian proposal
distributions for each parameter with widths chosen to give an
acceptance rate of 20 -- 60% of the steps.  See Figures
\ref{fig:hist13} and \ref{fig:hist24} for marginal distributions for
the final models.

The intent of MCMC is to explore the phase space, not to find one
optimal model.  Even the best model in an MCMC chain is not a good
replacement for the model found by a minimizer, because MCMC is
unlikely to land \textit{exactly} on the minimum that a minimizer
easily finds to machine precision.  If an MCMC chain finds a lower
\math{\chi\sp{2}} value than the minimizer's, then it has entered the
basin of attraction around a better local minimum, and a minimizer
will almost certainly find an even better \math{\chi\sp{2}} starting
from the MCMC's best value.  We thus refit at such points and then
restart our MCMC routine from the new minimizer solution.  The
\math{\chi\sp{2}} used in the information criteria described below
refers to the global minimum of a given dataset and not merely the
sampled minimum from MCMC.  Although the differences may appear to be
small, at the extreme precisions required for high-contrast photometry
and models with many parameters, parameter values can differ by a
significant fraction of \math{1\sigma} between the global and MCMC
minima, even for converged chains.

The MCMC routine ran an initial ``burn in'' of a least 10\sp{5}
iterations to forget the initial starting conditions, and then used
two million iterations to sample the phase space near the fit
solution.  To test for adequate sampling, we ran four independent MCMC
chains, three started away from the initial minimizer location, and
calculated the \citet{gelman:1992} statistic for each parameter.
These were all within 1% of unity, indicating the chains converged.
We initially fit each channel separately with all free model
parameters as MCMC jump parameters (see Table \ref{tab:ramp} in the
appendix).  Then we pair the channels observed together, fitting a
common eclipse phase and duration (see Table \ref{tab:fits}).  Due to
high correlations, the MCMC sampling becomes very inefficient with all
the parameters free in the joint fit.  Estimates of the interesting
parameters (eclipse depth, time, and duration) are unaffected if we
freeze \math{r\sb{2}} and the ingress and egress times at several
different values.  We set \math{r\sb{2}} from the independent
light-curve fits and the ingress and egress times as predicted by the
\citep{hebb:2009} orbit.

A recent re-analysis of older data by \citet{knutson:2009}
demonstrates that the complex models required to fit
\textit{Spitzer's} systematics can have multiple, comparable
\math{\chi\sp{2}} minima in different parts of phase space.  These
minima may change their relative depths given different systematic
models (e.g., exponential {\em vs.}\/ log-plus-linear ramps), resulting
in different conclusions.  To control for this, we fit data from a
range of photometry apertures with many combinations of analytic model
components (see Table \ref{tab:ramp} in the appendix) before choosing
Eq.\ \ref{eq:vip} -- Eq.\ \ref{eq:linramp}.  The models included
quadratic and logarithmic-plus-linear ramps and a variety of
polynomial intrapixel models.  Additionally, we drop a small number of
initial points to allow the pointing and instrument to stabilize,
which vastly improved the fits.

Choices among photometry apertures and numbers of dropped points are
choices between different datasets fit with the same models, so we
minimize the SDNR, removing the fewest points consistent with low
SDNR.  The model lines in Figure \ref{fig:lightcurves} show the
included points.

Once we have selected the dataset (by choice of aperture and dropped
points according to SDNR), we may apply any of several information
criteria to compare models with different numbers of free parameters
\citep{liddle:2007}.  These criteria have specific goals and
assumptions, so none is perfectly general, but two have broad
application.  The Akaike Information Criterion,
\begin{equation}
{\rm AIC} = \chi\sp{2} + 2k,
\end{equation}
where \math{k} is the number of free parameters, applies when the goal
is accurate prediction of future data; its derivation is valid even
when the candidate models might not include the theoretically correct
one (as is the case, so far, for {\em Spitzer}\/ intrapixel and ramp
modeling).  The Bayesian Information Criterion,
\begin{equation}
{\rm BIC} = \chi\sp{2} + k\ln N,
\end{equation}
where \math{N} is the number of data points, applies when the goal is
identifying the theoretically correct model, which is known to be one
of those being considered.  The best model minimizes the chosen
information criterion.  The ratio of probabilities favoring one model
over another is \math{\exp(\Delta{\rm BIC}/2)}, where \math{\Delta {\rm
    BIC}} is the difference in BIC between models, but the difference
in AIC between models has no simple calibration to a probability or
significance level.

These goals give different answers for finite datasets.  If the right
model is a candidate, the BIC will do better than AIC as the number of
points increases; if not, which is better depends on the sample size
and on how close the candidate models are to the (absent) correct
model.  Other information criteria exist, but are either tailored to
specific circumstances or are still being vetted by statisticians.
The criteria solve different problems, but the goal of a multi-model
analysis is not always easily classified as solely predictive or
explanatory, so there is some elasticity regarding the choice of an
appropriate criterion.

We calculate AIC and BIC for hundreds of models, and reject most of
them on this basis (see Table \ref{tab:ramp} in the appendix).  For the final
decision, we also consider the level of correlation in the residuals.
For this, we plot root-mean-squared (RMS) model residuals
\textit{vs.}\/ bin size (\citealp{pont:2006}, \citealp{winn:2008}) and
compare to the theoretical \math{1/\sqrt{N}} RMS scaling.  Figure
\ref{fig:noisecorr} demonstrates the lack of significant photometric
noise correlation in our final models.  In some cases, we prefer
less-correlated models with insignificantly poorer AIC or BIC (e.g.,
channel 1).  Differences in interesting parameter values (eclipse
depth, time, and duration) for such near-optimal alternatives are
\math{\lesssim1\sigma}.

Given the questions raised by re-analyses of certain {\em Spitzer}
exoplanet datasets \citep{knutson:2009, beaulieu:2010}, we consider it
critical that investigators disclose the details of their analyses
both so that readers can assess the quality of the analysis and so
that others may make meaningful comparisons in subsequent analyses of
the same data (e.g., did they find a better \math{\chi\sp{2}}?).  It
is important to include a full description of the centering,
photometry, uncertainty assessment, model fitting, correlation tests,
phase-space exploration, and convergence tests.  A listing of
alternative model fits and their quality may build confidence that
there is not a much better model than those tried.  One must identify
the particular \math{\chi\sp{2}} minimum explored by reporting even
nuisance parameter values, such as those in the intrapixel and ramp
curves.  

\begin{figure}[htb]
\if\submitms y \setcounter{fignum}{\value{figure}}
\addtocounter{fignum}{1} \newcommand\fignam{f\arabic{fignum}.eps}
\else \newcommand\fignam{figs/ch13-corr.eps} \fi
\includegraphics[width=\columnwidth, clip]{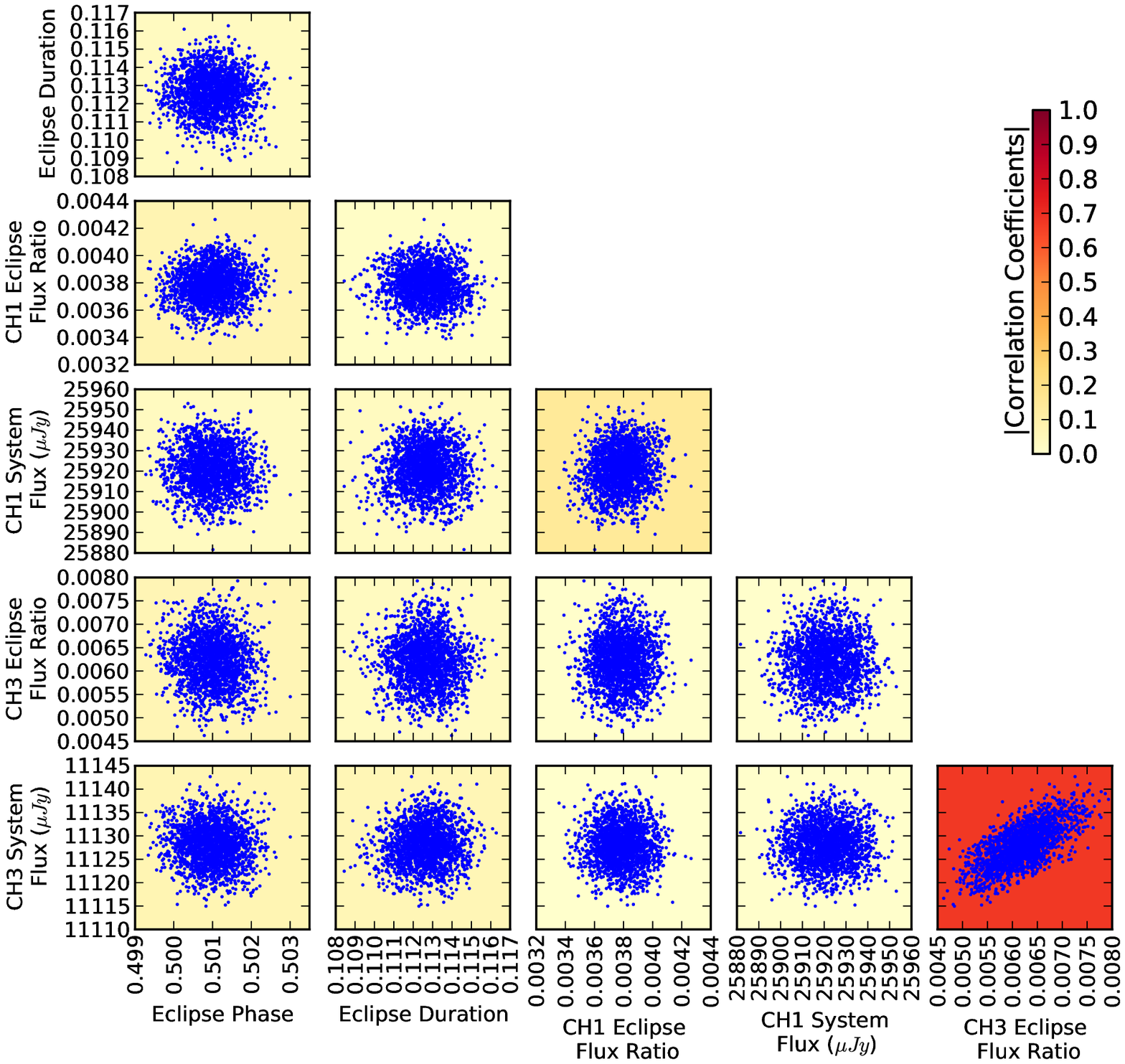}
\figcaption{\label{fig:corr13} Parameter correlations for 3.6 and 5.8
  {\microns}.  To decorrelate the Markov chains and unclutter the
  plot, one point appears for every 1000\sp{th} MCMC step.  Each panel
  contains all the points.}
\end{figure}

\begin{figure}[htb]
\if\submitms y \setcounter{fignum}{\value{figure}}
\addtocounter{fignum}{1} \newcommand\fignam{f\arabic{fignum}.eps}
\else \newcommand\fignam{figs/ch24-corr.eps} \fi
\includegraphics[width=\columnwidth, clip]{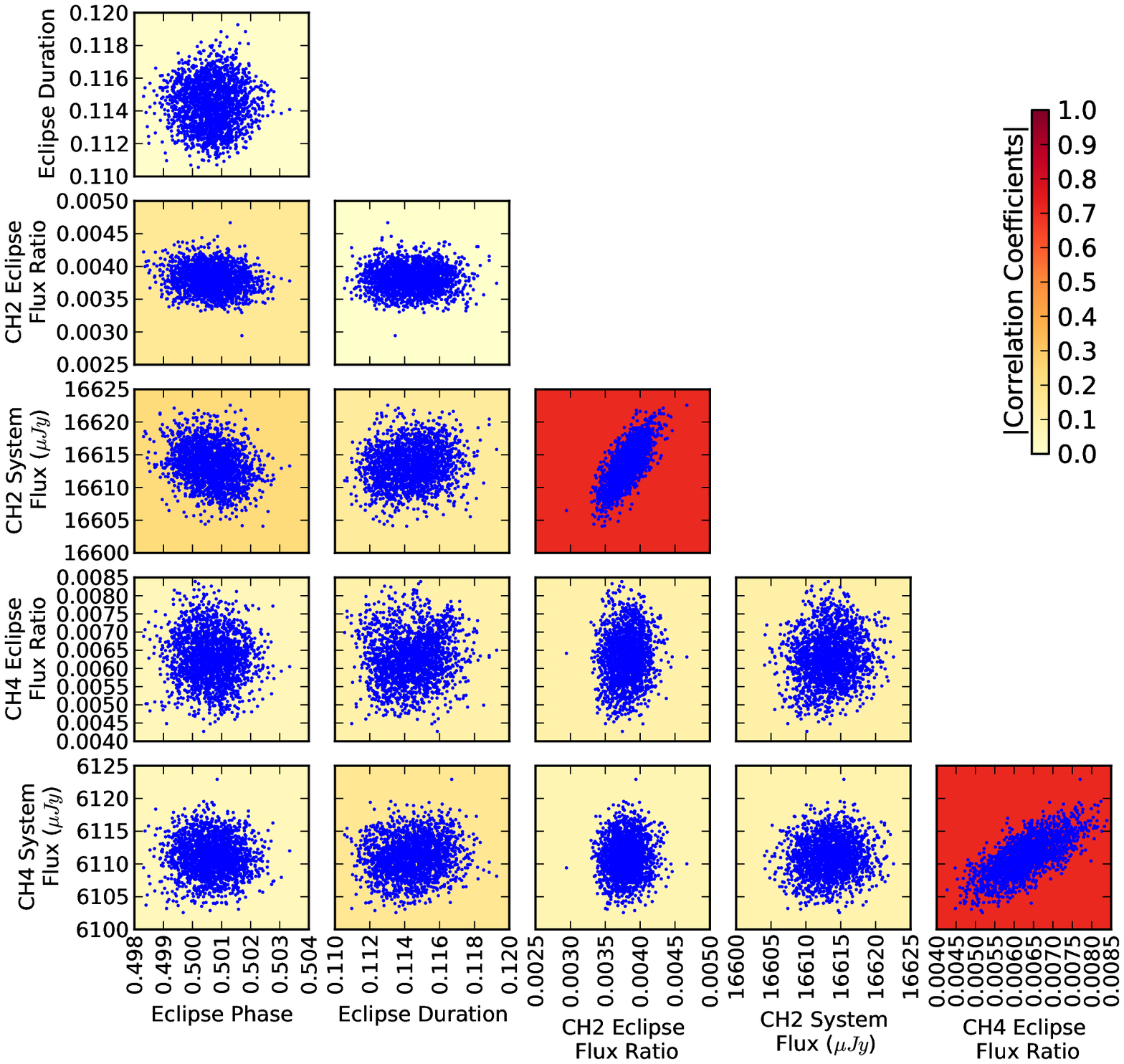}
\figcaption{\label{fig:corr24} Parameter correlations for 4.5 and 8.0
  {\microns}.  To decorrelate the Markov chains and unclutter the
  plot, one point appears for every 1000\sp{th} MCMC step.  Each panel
  contains all the points.}
\end{figure}

Finally, the marginal posterior distibutions (i.e., the parameter
histograms) and plots of their pairwise correlations help in assessing
whether the phase space minimum is global and in determining parameter
uncertainties.  We present these plots for the astrophysical
parameters in Figures \ref{fig:corr13}, \ref{fig:corr24},
\ref{fig:hist13}, & \ref{fig:hist24}.  The electronic supplement to
this article includes data files containing the photometry, best-fit
models, centering data, etc..  We encourage all investigators to make
similar disclosure in future reports of exoplanetary transits and
eclipses.

\section{DYNAMICS}
\label{sec:dyn}

\citet{hebb:2009} detect a non-zero eccentricity for WASP-12b that
should be observable in the timing of the secondary eclipse.  Our two
secondary eclipse phases (Table \ref{tab:fits}) are within
\math{2\sigma} of \math{\phi} = 0.5 for the \citet{hebb:2009}
ephemeris, and taken together imply \math{e\cos\omega} = 0.0016 {\pm}
0.0007.  This indicates that if the planet's orbit is eccentric, then
\math{\omega} is closely aligned with our line of sight.  Recognizing
the unlikelihood of this configuration (which implicitly questions the
\citealp{lopez:2009} eclipse phase), this section nonetheless
considers the possibility of significant eccentricity, with precession
between the \citet{lopez:2009} eclipse phase and \textit{Spitzer's}.
Subsequent to the initial submission of this paper, \citet{lopez:2010}
increased their uncertainty by a factor of three.  Since the arXiv
postings of both \citet{lopez:2009} and the submitted version of this
paper (arXiv 1003.2763v1) raised some community discussion, we now
treat both cases to explain how this adjustment changes our
conclusions.

\comment{ e cos omega:
((0.5012/0.0006**2+0.5007/0.0007**2)/(1/0.0006**2+1/.0007**2)-0.5)*np.pi/2.
0.0015523163700090366
((0.0006**2+0.0007**2)**0.5)/2*np.pi/2.
0.00072410132841189577
}

\if\submitms y
\clearpage
\fi
\atabon\begin{deluxetable}{lll}
\tablecaption{\label{tab:ttv} Transit Timing Data}
\tablewidth{0pt}
\tablehead{
\colhead{Mid-Transit Time (HJD)} &Uncertainty&Source\tablenotemark{a}}
\startdata
2453264.7594    &0.0048         &WASP Team \\
2454120.4290    &0.0070         &WASP Team \\
2454129.1600    &0.0017         &WASP Team \\
2454508.9761    &0.0002         &\citet{hebb:2009}\\
2454515.52464   &0.00016        &WASP Team \\
2454552.6218    &0.0034         &WASP Team \\
2454836.4026    &0.0006         &Veli-Pekka Hentunen, AXA\\
2454837.4955    &0.0013         &Alessandro Marchini, AXA\\
2454840.7704    &0.001          &Bruce Gary, AXA\\
2454848.41003   &0.00213        &Franti\u{s}ek Lomoz, TRESCA\\
2454860.41473   &0.0023         &Yenal \"{O}\u{g}men, TRESCA\\
2454860.4176    &0.00132        &Jaroslav Trnka, TRESCA\\
2454883.33312   &0.0056         &Alessandro Marchini, AXA\\
2454908.4372    &0.001          &Ramon Naves, AXA\\
2454931.35739   &0.00098        &Lubos Br\'{a}t, TRESCA\\
2455136.54322   &0.00066        &Leonard Kornos and\\
                &               &Peter Veres, TRESCA\\
2455151.82129   &0.00141        &Stan Shadick, TRESCA\\
2455164.92317   &0.00149        &Stan Shadick, TRESCA\\
2455172.5620    &0.00014        &Mikael Ingemyr, TRESCA\\
2455197.6628    &0.00203        &Brian Tieman, TRESCA\\
2455198.75595   &0.00141        &Brian Tieman, TRESCA\\
2455219.48996   &0.00131        &Lubos Br\'{a}t, TRESCA
\enddata
\tablenotetext{a}{The Amateur Exoplanet Archive (AXA,
  http://brucegary.net/AXA/x.htm) and TRansiting ExoplanetS and
  CAndidates group (TRESCA, http://var2.astro.cz/EN/tresca/index.php)
  supply their data to the Exoplanet Transit Database (ETD,
  http://var2.astro.cz/ETD/), which performs the uniform transit
  analysis described by \citet{poddany:2010}.  The ETD web site
  provided the AXA and TRESCA numbers in this table.}
\end{deluxetable}\ataboff
\if\submitms y
\clearpage
\fi
\placetable{tab:ttv}

\if\submitms y
\clearpage
\fi
\atabon\begin{deluxetable}{rr@{\,{\pm}\,}rr@{\,{\pm}\,}rr@{\,\,}}
\tablecaption{\label{tab:orbit} Orbital Fits}
\tablewidth{0pt}
\tablehead{
\colhead{Parameter} &
\mctc{No Precession} &
\mctc{With Precession}
}
\startdata
\comment{                                                 NO PRECESSION   ERR                        PRECESSION     ERR     }
\math{e \sin \omega\sb{0}\tablenotemark{a}}              &    -0.065     & 0.014                    &    -0.065    & 0.014   \\
\math{e \cos \omega\sb{0}\tablenotemark{a}}              &     0.0014    & 0.0007                   &    -0.0058   & 0.0027  \\
\math{e}                                                 &     0.065     & 0.014                    &     0.065    & 0.014   \\
\math{\omega\sb{0}} (\degree)                            &   -88.8       & 0.9                      &   -95.1      & 2.3     \\
\math{\dot{\omega}} (\degree d\sp{-1})\tablenotemark{a}  &     0         & 0                        &     0.026    & 0.009   \\
\math{P\sb{s}} (days)\tablenotemark{a}                   &     1.0914240 & 3\math{\times 10\sp{-7}} &     1.091436 & 4\math{\times 10\sp{-6}}  \\
\math{P\sb{a}} (days)                                    &     1.0914240 & 3\math{\times 10\sp{-7}} &     1.091521 & 3\math{\times 10\sp{-5}}  \\
\math{T\sb{0}} (MJD)\tablenotemark{a,b}                  &   508.97686   & 0.00012                  &   508.97686  & 0.00012 \\
\math{K} (ms\sp{-1})\tablenotemark{a}                    &   224         & 4                        &   224        & 4     \\
\math{\gamma} (ms\sp{-1})\tablenotemark{a}               & 19087         & 3                        & 19088        & 3     \\
BIC                                                      &  \mctc{101.0} & \mctc{97.6}                                             
\enddata
\tablenotetext{a}{MCMC Jump Parameter.}
\tablenotetext{b}{MJD = JD - 2,454,000.}
\end{deluxetable}\ataboff
\if\submitms y
\clearpage
\fi
\placetable{tab:orbit}

\if\submitms y
\clearpage
\fi
\if\submitms y
\clearpage
\fi

\if\submitms y
\clearpage
\fi
\if\submitms y
\clearpage
\fi

\if\submitms y
\clearpage
\fi
\if\submitms y
\clearpage
\fi

\if\submitms y
\clearpage
\fi
\if\submitms y
\clearpage
\fi

\if\submitms y
\clearpage
\fi
\begin{figure}[htb]
\if\submitms y
  \setcounter{fignum}{\value{figure}}
  \addtocounter{fignum}{1}
  \newcommand\fignama{f\arabic{fignum}a.eps}
  \newcommand\fignamb{f\arabic{fignum}b.eps}
  \newcommand\figwid{0.7}
\else
  \newcommand\fignama{figs/no-prec/oc.ps}
  \newcommand\fignamb{figs/prec/oc.ps}
  \newcommand\figwid{1.0}
\fi
\begin{center}
\includegraphics[width=\figwid\columnwidth, clip]{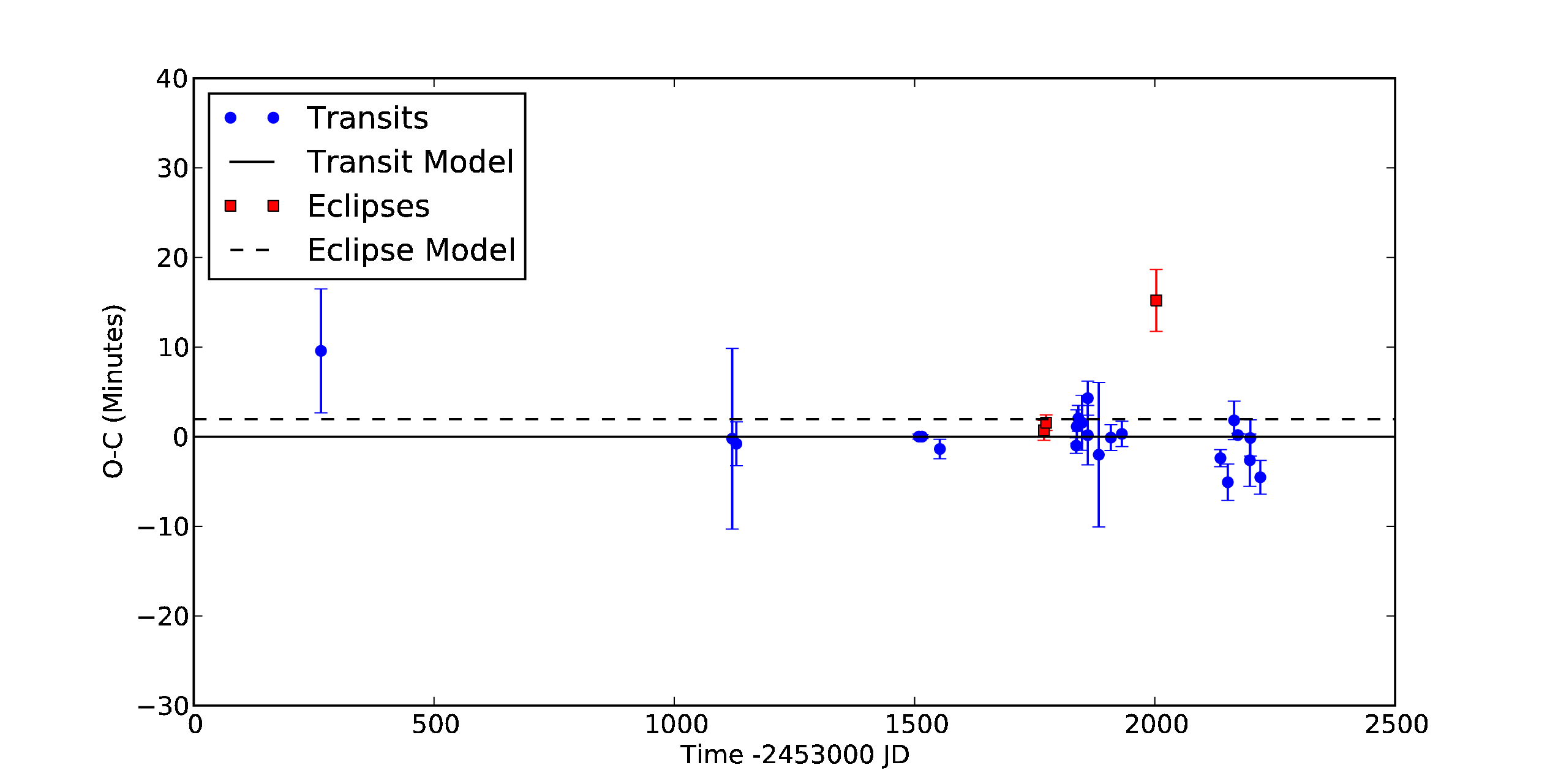}
\includegraphics[width=\figwid\columnwidth, clip]{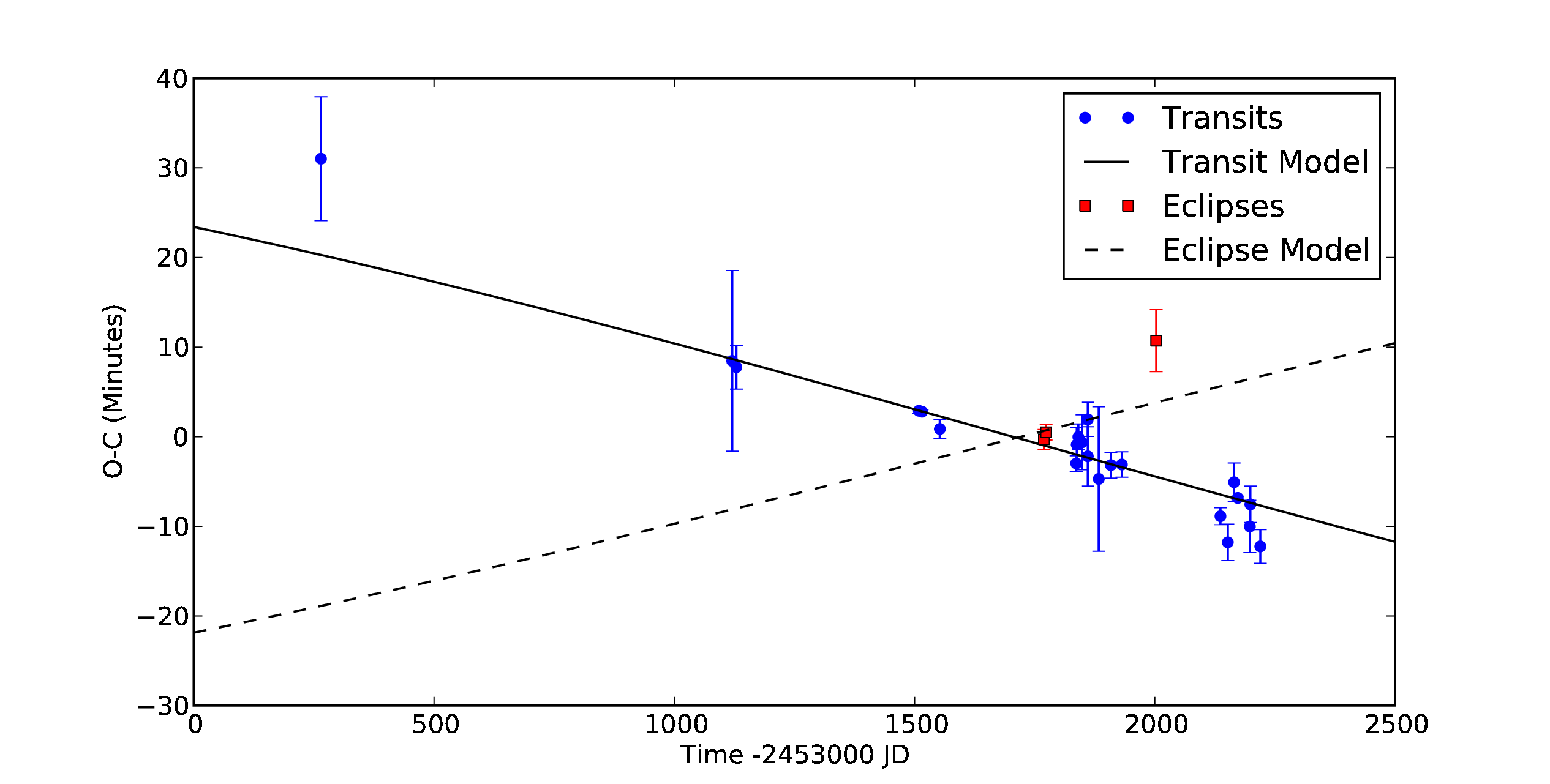}
\end{center}
\figcaption{\label{fig:prec} Transit times and orbit models of Table
  \ref{tab:orbit}.  Top: Non-precessing case.  Bottom: Precessing
  case.  Both diagrams show the difference (observed-minus-calculated,
  O-C) from a linear ephemeris determined from \math{P\sb{\rm s}} and
  \math{T\sb{0}} given for their respective cases in Table
  \ref{tab:orbit}.  This highlights deviations from the ephemeris.
  The dashed lines give eclipse times for the eccentric orbits of the
  fits.  In the non-precessing case, these lines are straight and
  horizontal, and the transits (which carry the most weight in
  determining the period) scatter about the line, as expected.
  However, the \citet{lopez:2009} point suggests a trend in the
  eclipses consistent with apsidal precession.  In the precessing
  case, the models are opposing sinusoids with a \sim32-minute
  amplitude and a 33-year period.  The curves cross approximately
  where \math{\omega = -90{\degrees}}.  Both curves and data were
  shifted upward by \math{-\frac{eP\sb{\rm{a}}}{\pi} \cos
    \omega\sb{0}} (about 2 minutes) to adjust for the modification in
  Eq.\ \ref{eq:gimenmod}, so the curves cross where O-C = 0.}
\end{figure}
\if\submitms y
\clearpage
\fi

We use an MCMC routine to fit a Keplerian model of the planet's orbit
to our secondary eclipse times, radial velocity data
\citep{hebb:2009}, transit timing data provided by the WASP team and
amateur observers (Table \ref{tab:ttv}), and the ground-based
secondary eclipse measurement of \citet{lopez:2009, lopez:2010}.
Because \citeauthor{lopez:2009} folded 1.5 complete eclipses, we
represent their point as a single observation taken during an orbit
halfway between their eclipses (HJD 2455002.8560 {\pm} 0.0024).  We
remove three in-transit radial velocity points due to
Rossiter-McLaughlin contamination, and correct the times of
mid-eclipse given in Table \ref{tab:fits} and by \citet{lopez:2009,
  lopez:2010} for light travel across the orbit by subtracting 22.8
seconds.  We note that eclipse observers should report uncorrected
times, as the correction depends on the orbit model and, in the
future, measurements may be uncertain at the level of model
uncertainty.

The amateur observers synchronize their clocks to within one second of
UTC by means such as Network Time Protocol (NTP) or radio signals from
atomic clocks.  In pre-publication discussions with
\citet{eastman:2010}, we determined that the amateurs' observing
software, MaximDL, did not account for leap seconds, nor did the
software of most of our professional contributors.  We thus made the
adjustment ourselves as needed.

In our model,
\begin{eqnarray}
\label{eq:chisq}
\chi\sp{2} = \sum{\left(\frac{v\sb{\rm{rv,o}}  -v\sb{\rm{rv,m}}}{\sigma\sb{\rm{rv}}}\right)}\sp{2}
\atab+\atab  \sum{\left(\frac{t\sb{\rm{tr,o}}  -t\sb{\rm{tr,m}}}{\sigma\sb{\rm{tr}}}\right)}\sp{2} \nonumber \\
\atab+\atab  \sum{\left(\frac{t\sb{\rm{ecl,o}} -t\sb{\rm{ecl,m}}}{\sigma\sb{\rm{ecl}}}\right)}\sp{2},
\end{eqnarray}
where \math{v\sb{\rm{rv,o}}}, \math{t\sb{\rm{tr,o}}},
\math{t\sb{\rm{ecl,o}}} are the observed radial velocities, transit
times, and eclipse times, respectively; \math{\sigma\sb{\rm{rv}}},
\math{\sigma\sb{\rm{tr}}}, \math{\sigma\sb{\rm{ecl}}} are their
respective uncertainties, and \math{v\sb{\rm{rv,m}}},
\math{t\sb{\rm{tr,m}}}, \math{t\sb{\rm{ecl,m}}} are the respective
model calculations.

Table \ref{tab:orbit} gives our best-fit results using the original
\citet{lopez:2009}; differences from our arXiv posting are due to the
time corrections and the use of a minimizer to find the true
\math{\chi\sp{2}} minimum.  The eccentricity of \math{e} = 0.065 {\pm}
0.014 may be high due to poor constraints on \math{e\sin\omega}.  Our
dynamical fits considered only the transit and eclipse times and did
not directly fit the light curves, which could additionally have
modeled variable eclipse and transit durations.

A significantly positive eccentricity implies either extremely low
tidal dissipation (e.g., \math{Q\sb{\rm{p}} \gtrsim 10\sp{8}}; a tidal
evolution model could give a better limit, \citealp{mardling:2007},
\citealp{levrard:2009}) or a perturber such as another planet.  In the
latter case, coupling between the two planets could potentially drain
energy and angular momentum from the outer orbit to the point where it
is not able to maintain a large eccentricity for WASP-12b
\citep[e.g.,][]{mardling:2007}.  Tidal dissipation of a non-zero
eccentricity could account for the inflated radius of WASP-12b.  If
the orbit is actually circular, the bloated size \citep{hebb:2009}
requires either an energy source or new interior models.

As noted above, the planet's proximity to its star must raise huge
tidal bulges \citep{ragozzine:2009} that significantly contribute to
an aspherical planetary gravitational potential.  This would induce
apsidal precession measurable over short timescales through transit
and eclipse timing variations.  The rate of precession is proportional
to the tidal Love number, {\ktwop}, which describes the concentration
of the planet's interior mass \citep{ragozzine:2009}.  A lower
{\ktwop} implies more central condensation, but {\ktwop} alone does
not define a unique density profile \citep{batygin:2009}.  A nominal
value of {\ktwop} = 0.3 yields precession of
\sim0.05\math{\degree}d\sp{-1} for the orbit of WASP-12b.  A precise
measurement of the precession rate will therefore constrain the
planet's internal structure, as long as the eccentricity is
significantly non-zero \citep{ragozzine:2009}.  Conversely, the
absence of observable precession limits the eccentricity.

We added a constant precession term, \math{\dot{\omega}}, to our
model, and took the inclination to be \sim90{\degrees}, as the timing
effects due to inclination should be negligible and the available
timing data cannot directly constrain this quantity.  With these
assumptions we modified Eq.\ 15 of \citet{gimenez:1995} such that
\begin{eqnarray}
\label{eq:gimenmod}
 T\sb{\rm{tr}}  =  T\sb{\rm{0}} + P\sb{\rm{s}}E - \frac{eP\sb{\rm{a}}}{\pi} 
 \left( \cos\omega\sb{\rm{tr}} - \cos\omega\sb0 \right)],
\end{eqnarray}
where \math{T\sb{\rm{tr}}} is the time of mid-transit, \math{T\sb0} is
the transit time at orbit zero, \math{P\sb{\rm{s}}} is the sidereal
period, and \math{P\sb{\rm{a}}} is the anomalistic period, or time
between successive periastron passages.  The right bracket indicates
truncation of a series.  Furthermore,
\math{P\sb{\rm{a}}} is related to \math{P\sb{s}} by
\begin{eqnarray}
 P\sb{a} = \frac{P\sb{s}}{1 - P\sb{s}\frac{\dot{\omega}}{2\pi}}.
\end{eqnarray}
\math{E} is the number of elapsed sidereal periods since
\math{T\sb{0}} and \math{\omega\sb{tr} = \dot{\omega}(T\sb{tr} -
  T\sb{0}) + \omega\sb{0}}, where \math{\omega\sb{0}} is \math{\omega}
at \math{T\sb{0}}.  We expand the equation to fifth order in \math{e}
and solve iteratively for \math{T\sb{tr}}.  We compute the eclipse
time as a function of \math{e}, \math{\omega\sb{tr}}, \math{P\sb{a}},
and \math{T\sb{tr}}; radial velocity is computed as a function of
\math{\omega(t)}.

Fitting this model to the data with the \citet{lopez:2009} point, we
found that \math{\dot{\omega}} = 0.026 {\pm} 0.009{\degrees}d\sp{-1}, a
3\math{\sigma} result.  This corresponds to a precession period of 33
{\pm} 13 years and implies that \math{k\sb{2p}} = 0.15 {\pm} 0.08 (see
Table \ref{tab:orbit}).  This result depended on an unlikely alignment
of the orbit with our line of sight during the {\em Spitzer}\/
observations.  The revised \citet{lopez:2010} uncertainty dashed hopes
for detecting precession, however, as the model fit with that point
(Table \ref{tab:orbit2}) yields a marginal precession, and BIC prefers
the non-precessing case.  Even if the 4\math{\sigma} eccentricity
stands, measurement of precession awaits a longer observational
baseline.

\section{CONCLUSIONS}
\label{sec:concl}

The timing of the {\em Spitzer}\/ eclipses is consistent with a
circular orbit, and our best fit, including RV data and transit and
eclipse times, does not detect precession.

Although the \citet{lopez:2010} eclipse phase is now marginally
consistent with zero eccentricity, we note that this 0.9-{\micron}
observation could be affected by a wavelength-dependent asymmetry in
the planet's surface-brightness distribution that manifests itself as
a timing offset \citep{knutson:2007}.  This offset has a maximum
possible value of \math{R\sb{\rm{p}}/v\sb{\rm{p}} \approx 9} minutes,
where \math{v\sb{\rm p}} is the planet's orbital velocity.  This is
somewhat smaller than the observed variation in eclipse timing between
\citet{lopez:2010} and \textit{Spitzer}.

While we have not yet measured precession, the possible prolateness
should be measurable in high-accuracy, infrared transits and eclipses
\citep{ragozzine:2009}, such as we expect will be available from the
James Webb Space Telescope.  This would provide another constraint on
interior structure, one that does not depend on an elliptical orbit.

As this paper was in late stages of revision, \citet{croll:2010}
published three ground-based secondary eclipses and
\citet{husnoo:2010} produced additional radial velocity data.  These
datasets are consistent with a circular orbit for WASP-12b.

As the quality of these data attests (the signal-to-noise ratio of the
eclipse depth in channel 1 is over 29, second only to that for HD
189733b), WASP-12b has emerged as a highly observable
exoplanet. \citet{madhu:2010} report our analysis of the planet's
atmospheric composition.  Its phase curves, already in {\em
  Spitzer's}\/ queue, will enable the first observational discussion
of atmospheric dynamics on a prolate planet.

\if\submitms y
\clearpage
\fi
\atabon\begin{deluxetable}{rr@{\,{\pm}\,}rr@{\,{\pm}\,}rr@{\,\,}}
\tablecaption{\label{tab:orbit2} Revised Orbital Fits}
\tablewidth{0pt}
\tablehead{
\colhead{Parameter} &
\mctc{No Precession} &
\mctc{With Precession}
}
\startdata
\comment{                                                    NO PRECESSION   ERR                        PRECESSION     ERR     }
\math{e \sin \omega\sb{0}\tablenotemark{a} }              &    -0.063     & 0.014                    &    -0.065    & 0.015   \\
\math{e \cos \omega\sb{0}\tablenotemark{a} }              &     0.0011    & 0.00072                  &    -0.0036   & 0.0045  \\
\math{e}                                                  &     0.063     & 0.014                    &     0.065    & 0.015   \\
\math{\omega\sb{0}} (\degree)                             &   -89.0       & 0.8                      &    -93       & 5     \\
\math{\dot{\omega}} (\degree d\sp{-1}\tablenotemark{a})   &     0         & 0                        &     0.017    & 0.019  \\
\math{P\sb{s}} (days) \tablenotemark{a}                   &     1.0914240 & 3\math{\times 10\sp{-7}} &    1.0914315 & 7\math{\times 10\sp{-6}}  \\
\math{P\sb{a}} (days)                                     &     1.0914240 & 3\math{\times 10\sp{-7}} &    1.0914872 & 7\math{\times 10\sp{-5}}  \\
\math{T\sb{0}} (MJD)\tablenotemark{a,b}                   &   508.97683   & 0.00012                  &   508.97685  & 0.00012 \\
\math{K} (ms\sp{-1}\tablenotemark{a})                     &   225         & 4                        &   224        & 4     \\
\math{\gamma} (ms\sp{-1}\tablenotemark{a})                & 19087         & 3                        & 19088        & 3     \\
BIC                                                       &  \mctc{90.1}  & \mctc{92.8}                                             
\enddata
\tablenotetext{a}{MCMC Jump Parameter.}
\tablenotetext{b}{MJD = JD - 2,454,000.}
\end{deluxetable}\ataboff
\if\submitms y
\clearpage
\fi
\placetable{tab:orbit2}

\acknowledgments We thank the observers listed in Table \ref{tab:ttv}
for allowing us to use their results, and the organizers of the
Exoplanet Transit Database for coordinating the collection and uniform
analysis of these data.  The IRAC data are based on observations made
with the \textit{Spitzer} Space Telescope, which is operated by the
Jet Propulsion Laboratory, California Institute of Technology under a
contract with NASA.  Support for this work was provided by NASA
through an award issued by JPL/Caltech.  The point of M.\ Ingemyr is
based on observations made with the Nordic Optical Telescope (NOT),
operated on the island of La Palma jointly by Denmark, Finland,
Iceland, Norway, and Sweden, in the Spanish Observatorio del Roque de
los Muchachos Instituto Astrofisica de Canarias, and ALFOSC, which is
owned by the Instituto Astrofisica de Andalucia (IAA) and operated at
the NOT under agreement between IAA and NBlfAFG of the Astronomical
Observatory of Copenhagen.  We thank contributors to SciPy,
Matplotlib, and the Python Programming Language, W.\ Landsman and
other contributors to the Interactive Data Language Astronomy Library,
the free and open-source community, the NASA Astrophysics Data System,
and the JPL Solar System Dynamics group for free software and
services.

\appendix

\if\submitms n
\LongTables
\fi
\if\submitms y
\clearpage
\fi
\atabon\begin{deluxetable}{lccccc}
\if\submitms y
\tabletypesize{\tiny}
\fi
\tablecaption{\label{tab:ramp} Candidate Models}
\tablewidth{0pc}
\tablehead{
\colhead{Model} &
\colhead{Ap\tablenotemark{a}} &
\colhead{NFP\tablenotemark{b}} &
\colhead{BIC\tablenotemark{c}} &
\colhead{AIC\tablenotemark{c}} &
\colhead{SDNR\tablenotemark{c}} 
}
\startdata
\multicolumn{6}{l}{Channel 1, All Intrapixel Parameters Free:}\\
\\
\multicolumn{6}{l}{1553 points, uncertainties multiplied by 0.30946} \\
Linear     &  3.00  &  11  &  1622.8  &  1564.0  &  0.00232818\\
Quadratic  &  3.00  &  11  &  1612.1  &  1553.3  &  0.00231926\\
Log+Linear &  3.00  &  12  &  1624.5  &  1560.3  &  0.00232366\\
Rising Exp &  3.00  &  10  &  1611.8  &  1558.3  &  0.00232512\\
\\
\multicolumn{6}{l}{1544 points, uncertainties multiplied by 0.31028} \\
Linear     &  3.25  &  11  &  1613.8  &  1555.0  &  0.00230984\\
Quadratic  &  3.25  &  11  &  1602.7  &  1543.9  &  0.00230068\\
Log+Linear &  3.25  &  12  &  1615.1  &  1551.0  &  0.00230511\\
Rising Exp &  3.25  &  10  &  1602.4  &  1549.0  &  0.00230655\\
\\
\multicolumn{6}{l}{1536 points, uncertainties multiplied by 0.31130} \\
Linear     &  3.50  &  11  &  1605.7  &  1547.0  &  0.00229716\\
Quadratic  &  3.50  &  11  &  1591.8  &  1533.0  &  0.00228574\\
Log+Linear &  3.50  &  12  &  1605.9  &  1541.9  &  0.00229146\\
Rising Exp &  3.50  &  10  &  1592.3  &  1539.0  &  0.00229226\\
\\
\multicolumn{6}{l}{1532 points, uncertainties multiplied by 0.31286} \\
Linear     &  3.75  &  11  &  1601.7  &  1543.0  &  0.00229111\\
Quadratic  &  3.75  &  11  &  1588.9  &  1530.2  &  0.00228061\\
Log+Linear &  3.75  &  12  &  1601.5  &  1537.5  &  0.00228509\\
Rising Exp &  3.75  &  10  &  1588.8  &  1535.4  &  0.00228651\\
\\
\multicolumn{6}{l}{1530 points, uncertainties multiplied by 0.31547} \\
Linear     &  4.00  &  11  &  1599.7  &  1541.0  &  0.00229468\\
Quadratic  &  4.00  &  11  &  1586.5  &  1527.8  &  0.00228382\\
Log+Linear &  4.00  &  12  &  1599.2  &  1535.2  &  0.00228842\\
Rising Exp &  4.00  &  10  &  1586.4  &  1533.0  &  0.00228974\\
\\
\multicolumn{6}{l}{Channel 1, Intrapixel with only \math{x\sp{2}} and
  \math{y} free:}\\
\\
\multicolumn{6}{l}{1553 points, uncertainties multiplied by 0.30916} \\
Linear     &  3.00  &  8   &  1603.8  &  1561.0  &  0.00232820\\
Quadratic  &  3.00  &  8   &  1593.0  &  1550.2  &  0.00231930\\
Log+Linear &  3.00  &  9   &  1605.0  &  1556.9  &  0.00232329\\
Rising Exp &  3.00  &  7   &  1592.7  &  1555.3  &  0.00232514\\
\\                                                            
\multicolumn{6}{l}{1544 points, uncertainties multiplied by 0.31013} \\
Linear     &  3.25  &  8   &  1594.7  &  1552.0  &  0.00231106\\
Quadratic  &  3.25  &  8   &  1583.9  &  1541.2  &  0.00230213\\
Log+Linear &  3.25  &  9   &  1596.1  &  1548.1  &  0.00230627\\
Rising Exp &  3.25  &  7   &  1583.4  &  1546.0  &  0.00230874\\
\\                                                            
\multicolumn{6}{l}{1536 points, uncertainties multiplied by 0.31106} \\
Linear     &  3.50  &  8   &  1586.7  &  1544.0  &  0.00229865\\
Quadratic  &  3.50  &  8   &  1572.6  &  1529.9  &  0.00228619\\
Log+Linear &  3.50  &  9   &  1587.0  &  1539.0  &  0.00229298\\
Rising Exp &  3.50  &  7   &  1573.2  &  1535.8  &  0.00229367\\
\\                                                            
\multicolumn{6}{l}{1532 points, uncertainties multiplied by 0.31261} \\
Linear     &  3.75  &  8   &  1582.7  &  1540.0  &  0.00229250\\
Quadratic  &  3.75  &  8   &  1569.7  &  1527.0  &  0.00228188\\
Log+Linear &  3.75  &  9   &  1583.3  &  1535.3  &  0.00228609\\
Rising Exp &  3.75  &  7   &  1569.6  &  1532.3  &  0.00228781\\
\\                                                            
\multicolumn{6}{l}{1530 points, uncertainties multiplied by 0.31521} \\
Linear     &  4.00  &  8   &  1580.7  &  1538.0  &  0.00229595\\
Quadratic  &  4.00  &  8   &  1567.3  &  1524.6  &  0.00228496\\
Log+Linear &  4.00  &  9   &  1583.8  &  1535.8  &  0.00229263\\
Rising Exp &  4.00  &  7   &  1567.2  &  1529.9  &  0.00229094\\
\\
\multicolumn{6}{l}{Channel 2, All Intrapixel Parameters Free:} \\
\\
\multicolumn{6}{l}{1465 points, uncertainties multiplied by 0.44456} \\
No Ramp     &  3.50  &   9  &  1521.6  &  1474.0  &  0.00326693\\
Linear      &  3.50  &  10  &  1501.7  &  1448.8  &  0.00323655\\
Quadratic   &  3.50  &  11  &  1506.8  &  1448.6  &  0.00323340\\
Falling Exp &  3.50  &  11  &  1509.9  &  1451.7  &  0.00323775\\
\\
\multicolumn{6}{l}{1460 points, uncertainties multiplied by 0.44663} \\
No Ramp     &  3.75  &   9  &  1516.6  &  1469.0  &  0.00326495\\
Linear      &  3.75  &  10  &  1496.9  &  1444.1  &  0.00323477\\
Quadratic   &  3.75  &  11  &  1502.3  &  1444.1  &  0.00323189\\
Falling Exp &  3.75  &  11  &  1505.1  &  1446.9  &  0.00323590\\
\\
\multicolumn{6}{l}{1457 points, uncertainties multiplied by 0.44875} \\
No Ramp     &  4.00  &   9  &  1513.6  &  1466.0  &  0.00326355\\
Linear      &  4.00  &  10  &  1492.6  &  1439.7  &  0.00323181\\
Quadratic   &  4.00  &  11  &  1497.7  &  1439.6  &  0.00322873\\
Falling Exp &  4.00  &  11  &  1500.8  &  1442.7  &  0.00323302\\
\\
\multicolumn{6}{l}{1449 points, uncertainties multiplied by 0.45211} \\
No Ramp     &  4.25  &   9  &  1505.5  &  1458.0  &  0.00327075\\
Linear      &  4.25  &  10  &  1483.7  &  1430.9  &  0.00323780\\
Quadratic   &  4.25  &  11  &  1488.5  &  1430.4  &  0.00323425\\
Falling Exp &  4.25  &  11  &  1492.0  &  1434.0  &  0.00323913\\
\\
\multicolumn{6}{l}{1435 points, uncertainties multiplied by 0.45715} \\
No Ramp     &  4.50  &   9  &  1491.4  &  1444.0  &  0.00329102\\
Linear      &  4.50  &  10  &  1470.1  &  1417.4  &  0.00325808\\
Quadratic   &  4.50  &  11  &  1474.8  &  1416.9  &  0.00325441\\
Falling Exp &  4.50  &  11  &  1478.4  &  1420.5  &  0.00325944\\
\\
\multicolumn{6}{l}{Channel 2, Intrapixel With Only \math{y\sp{2}} Free:} \\
\\
\multicolumn{6}{l}{1465 points, uncertainties multiplied by 0.44695} \\
No Ramp     &  3.50  &  5  &  1496.4  &  1470.0  &  0.00328991\\
Linear      &  3.50  &  6  &  1464.4  &  1432.6  &  0.00324507\\
Quadratic   &  3.50  &  7  &  1470.0  &  1433.0  &  0.00324266\\
Falling Exp &  3.50  &  7  &  1472.6  &  1435.6  &  0.00324626\\
\\
\multicolumn{6}{l}{1460 points, uncertainties multiplied by 0.44969} \\
No Ramp     &  3.75  &  5  &  1491.4  &  1465.0  &  0.00329263\\
Linear      &  3.75  &  6  &  1455.3  &  1423.6  &  0.00324297\\
Quadratic   &  3.75  &  7  &  1461.2  &  1424.2  &  0.00324081\\
Falling Exp &  3.75  &  7  &  1463.5  &  1426.5  &  0.00324417\\
\\
\multicolumn{6}{l}{1457 points, uncertainties multiplied by 0.45194} \\
No Ramp     &  4.00  &  5  &  1488.4  &  1462.0  &  0.00329209\\
Linear      &  4.00  &  6  &  1450.5  &  1418.8  &  0.00324027\\
Quadratic   &  4.00  &  7  &  1456.2  &  1419.2  &  0.00323791\\
Falling Exp &  4.00  &  7  &  1458.7  &  1421.8  &  0.00324155\\
\\
\multicolumn{6}{l}{1449 points, uncertainties multiplied by 0.45480} \\
No Ramp     &  4.25  &  5  &  1480.4  &  1454.0  &  0.00329562\\
Linear      &  4.25  &  6  &  1444.5  &  1412.9  &  0.00324584\\
Quadratic   &  4.25  &  7  &  1449.8  &  1412.9  &  0.00324293\\
Falling Exp &  4.25  &  7  &  1452.9  &  1415.9  &  0.00324720\\
\\
\multicolumn{6}{l}{1435 points, uncertainties multiplied by 0.45965} \\
No Ramp     &  4.50  &  5  &  1466.3  &  1440.0  &  0.00331453\\
Linear      &  4.50  &  6  &  1432.7  &  1401.1  &  0.00326654\\
Quadratic   &  4.50  &  7  &  1437.9  &  1401.0  &  0.00326347\\
Falling Exp &  4.50  &  7  &  1441.0  &  1404.1  &  0.00326791\\
\\
\multicolumn{6}{l}{Channel 3:}\\
\\
\multicolumn{6}{l}{1544 points, uncertainties multiplied by 0.91447} \\
No Ramp     &  2.50  &  2  &  1556.7  &  1546.0  &  0.01066331\\
Linear      &  2.50  &  3  &  1553.7  &  1537.6  &  0.01062468\\
Falling Exp &  2.50  &  4  &  1563.2  &  1541.9  &  0.01063552\\
\\
\multicolumn{6}{l}{1543 points, uncertainties multiplied by 0.92247} \\
No Ramp     &  2.75  &  2  &  1555.7  &  1545.0  &  0.01063659\\
Linear      &  2.75  &  3  &  1549.9  &  1533.9  &  0.01058879\\
Falling Exp &  2.75  &  4  &  1558.7  &  1537.4  &  0.01059726\\
\\
\multicolumn{6}{l}{1535 points, uncertainties multiplied by 0.93860} \\
No Ramp     &  3.00  &  2  &  1547.7  &  1537.0  &  0.01080977\\
Linear      &  3.00  &  3  &  1543.1  &  1527.1  &  0.01076529\\
Falling Exp &  3.00  &  4  &  1551.7  &  1530.4  &  0.01077319\\
\\
\multicolumn{6}{l}{1529 points, uncertainties multiplied by 0.95222} \\
No Ramp     &  3.25  &  2  &  1541.7  &  1531.0  &  0.01103641\\
Linear      &  3.25  &  3  &  1538.6  &  1522.6  &  0.01099631\\
Falling Exp &  3.25  &  4  &  1547.6  &  1526.2  &  0.01100564\\
\\
\multicolumn{6}{l}{1524 points, uncertainties multiplied by 0.96551} \\
No Ramp     &  3.50  &  2  &  1536.7  &  1526.0  &  0.01132113\\
Linear      &  3.50  &  3  &  1535.2  &  1519.2  &  0.01128630\\
Falling Exp &  3.50  &  4  &  1544.0  &  1522.7  &  0.01129501\\
\\
\multicolumn{6}{l}{1522 points, uncertainties multiplied by 0.97776} \\
No Ramp     &  3.75  &  2  &  1534.7  &  1524.0  &  0.01163603\\
Linear      &  3.75  &  3  &  1533.8  &  1517.8  &  0.01160264\\
Falling Exp &  3.75  &  4  &  1542.3  &  1521.0  &  0.01161056\\
\\
\multicolumn{6}{l}{1521 points, uncertainties multiplied by 0.99239} \\
No Ramp     &  4.00  &  2  &  1533.7  &  1523.0  &  0.01201476\\
Linear      &  4.00  &  3  &  1533.2  &  1517.2  &  0.01198222\\
Falling Exp &  4.00  &  4  &  1541.6  &  1520.3  &  0.01198978\\
\\
\multicolumn{6}{l}{Channel 4:}\\
\\
\multicolumn{6}{l}{1467 points, uncertainties multiplied by 0.64001} \\
No Ramp     &  2.00  &  2  &  1479.6  &  1469.0  &  0.01269320\\
Linear      &  2.00  &  3  &  1470.6  &  1454.8  &  0.01262427\\
Rising Exp  &  2.00  &  4  &  1477.1  &  1455.9  &  0.01261620\\
Log+Linear  &  2.00  &  5  &  1484.4  &  1458.0  &  0.01261728\\
Quadratic   &  2.00  &  4  &  1476.2  &  1455.0  &  0.01260762\\
\\
\multicolumn{6}{l}{1467 points, uncertainties multiplied by 0.65283} \\
No Ramp     &  2.25  &  2  &  1479.6  &  1469.0  &  0.01291715\\
Linear      &  2.25  &  3  &  1466.9  &  1451.1  &  0.01283127\\
Rising Exp  &  2.25  &  4  &  1473.8  &  1452.6  &  0.01282649\\
Log+Linear  &  2.25  &  5  &  1479.7  &  1453.2  &  0.01281869\\
Quadratic   &  2.25  &  4  &  1473.1  &  1452.0  &  0.01281909\\
\\
\multicolumn{6}{l}{1464 points, uncertainties multiplied by 0.67232} \\
No Ramp     &  2.50  &  2  &  1476.6  &  1466.0  &  0.01333255\\
Linear      &  2.50  &  3  &  1467.5  &  1451.6  &  0.01326022\\
Rising Exp  &  2.50  &  4  &  1474.6  &  1453.4  &  0.01325764\\
Log+Linear  &  2.50  &  5  &  1481.6  &  1455.1  &  0.01325392\\
Quadratic   &  2.50  &  4  &  1474.3  &  1453.1  &  0.01325268\\
\\
\multicolumn{6}{l}{1455 points, uncertainties multiplied by 0.68250} \\
No Ramp     &  2.75  &  2  &  1467.6  &  1457.0  &  0.01361319\\
Linear      &  2.75  &  3  &  1459.1  &  1443.2  &  0.01354198\\
Rising Exp  &  2.75  &  4  &  1466.4  &  1445.2  &  0.01354054\\
Log+Linear  &  2.75  &  5  &  1474.2  &  1447.8  &  0.01354051\\
Quadratic   &  2.75  &  4  &  1466.3  &  1445.2  &  0.01353948\\
\\
\multicolumn{6}{l}{1448 points, uncertainties multiplied by 0.69698} \\
No Ramp     &  3.00  &  2  &  1460.6  &  1450.0  &  0.01404468\\
Linear      &  3.00  &  3  &  1451.3  &  1435.4  &  0.01396745\\
Rising Exp  &  3.00  &  4  &  1458.5  &  1437.4  &  0.01396580\\
Log+Linear  &  3.00  &  5  &  1467.0  &  1440.6  &  0.01396978\\
Quadratic   &  3.00  &  4  &  1458.5  &  1437.3  &  0.01396422\\
\\
\multicolumn{6}{l}{1443 points, uncertainties multiplied by 0.71158} \\
No Ramp     &  3.25  &  2  &  1455.5  &  1445.0  &  0.01457611\\
Linear      &  3.25  &  3  &  1443.8  &  1428.0  &  0.01448370\\
Rising Exp  &  3.25  &  4  &  1451.0  &  1429.9  &  0.01448177\\
Log+Linear  &  3.25  &  5  &  1456.9  &  1430.6  &  0.01448402\\
Quadratic   &  3.25  &  4  &  1450.9  &  1429.8  &  0.01447941\\
\\
\multicolumn{6}{l}{1440 points, uncertainties multiplied by 0.72888} \\
No Ramp     &  3.50  &  2  &  1452.5  &  1442.0  &  0.01528549\\
Linear      &  3.50  &  3  &  1437.9  &  1422.1  &  0.01517316\\
Rising Exp  &  3.50  &  4  &  1445.1  &  1424.0  &  0.01517067\\
Log+Linear  &  3.50  &  5  &  1451.3  &  1424.9  &  0.01517428\\
Quadratic   &  3.50  &  4  &  1444.9  &  1423.8  &  0.01516664\\ 
\\
\multicolumn{6}{l}{1431 points, uncertainties multiplied by 0.75781} \\
No Ramp     &  4.00  &  2  &  1443.5  &  1433.0  &  0.01694407\\
Linear      &  4.00  &  3  &  1424.0  &  1408.2  &  0.01679042\\
Rising Exp  &  4.00  &  4  &  1431.1  &  1410.1  &  0.01678704\\
Log+Linear  &  4.00  &  5  &  1437.9  &  1411.6  &  0.01679340\\
Quadratic   &  4.00  &  4  &  1431.0  &  1409.9  &  0.01678180\\

\enddata
\tablenotetext{a}{Aperture radius in pixels}
\tablenotetext{b}{Number of free parameters (\math{k} in the text)}
\tablenotetext{c}{Compare between the aperture sizes, for the same
  model, by SDNR.  Compare within the aperture sizes by BIC and AIC.}
\end{deluxetable}\ataboff
\if\submitms y
\clearpage
\fi
\placetable{tab:ramp}

\nojoe{
\if\submitms y
  \newcommand\bblnam{ms}
\else
  \newcommand\bblnam{wasp12bdyn}
\fi
\bibliography{\bblnam}}

\end{document}